# eUDEVS: Executable UML with DEVS Theory of Modeling and Simulation

**José L. Risco-Martín[1](*), Saurabh Mittal[2], Jesús M. de la Cruz[1] and Bernard P. Zeigler[3]**
(*) Corresponding author
[1]Departamento de Arquitectura de Computadores y Automática
Facultad de Informática, Universidad Complutense de Madrid
C/Prof. José García Santesmases, s/n
28040 Madrid, Spain
Tel: (+34) 91 394 7603
{jlrisco, jmcruz}@dacya.ucm.es

[2]Dunip Technologies,
New Delhi, India
saurabh.mittal@duniptechnologies.com

[3]Arizona Center for Integrative Modeling and Simulation
Electrical and Computer Engineering, University of Arizona
Tucson, AZ USA
zeigler@ece.arizona.edu

**Abstract**

*Modeling and Simulation (M&S)* for system design and prototyping is practiced today both in the industry and academia. M&S are two different areas altogether and have specific objectives. However, most of the times these two separate areas are taken together. The developed code is tightly woven around both the model and the underlying simulator that executes it. This constraints both the model development and the simulation engine that impacts scalability of the developed code. Furthermore, a lot of time is spent in development of a model because it needs both domain knowledge and simulation techniques, which also requires communication among users and developers. *Unified Modeling Language (UML)* is widely accepted in the industry, whereas *Discrete Event Specification (DEVS)* based modeling that separates the model and the simulator, provides a cleaner methodology to develop models and is much used in academia. DEVS today is used by engineers who understand discrete event modeling at a much detailed level and are able to translate requirements to DEVS modeling code. There have been earlier efforts to integrate UML and DEVS but they haven't succeeded in providing a transformation mechanism due to inherent differences in these two modeling paradigms. This paper presents an integrated approach towards crosstransformations between UML and DEVS using the proposed eUDEVS, which stands for executable UML based on DEVS. Further, we will also show that the obtained DEVS models belong to a specific class of DEVS models called Finite Deterministic DEVS (FD-DEVS) that is available as a W3C XML Schema in XFD-DEVS. We also put the proposed eUDEVS in a much larger unifying framework called DEVS Unified Process that allows bifurcated model-continuity based lifecycle methodology for systems M&S. Finally, we demonstrate the laid concepts with a complete example.

## 1. Introduction

It is difficult to develop a simulation model in the early phase of system development since it requires a high level knowledge of these three aspects: modeling techniques, the system domain and the model execution paradigms. The model development platform may be completely different from the model execution platform but most of the time both are treated as one. The developed model is so much customized to the problem at hand that it impedes extensibility on the said aspects. It becomes necessary to answer questions about properties (most notably behavior) of the whole system [1]. It requires intensive cooperation among domain experts and modeling experts as both demand totally different expertise. The entire system M&S process starts with elicitation of system requirements and ultimately translating them into an executable modeling code. The simple act of executing the model is termed as *simulation*. In addition, the *Department of Defense (DoD)* has strongly recommended applying M&S techniques to validate the requirements during the system development [2]. We need a practical and efficient way of applying

*Modeling and Simulation (M&S)* to the development of system under design in early phases of system development. Most importantly, we must separate the art of modeling with the model platform so that the subject experts can focus on the model abstraction rather than the modeling platform. In other words, a platform independent model (PIM) is the preferred way that would aid the subject matter expert to participate in the modeling process directly.

*Unified Modeling Language (UML)* is one of the preferred means of communication between the domain experts and modeling experts. UML is very powerful in terms of its graphical representation but diminishes in quality when it comes to execution of the UML model. Executable UML [3] is a working draft and is not operational in its current state today. We propose a design flow and a set of transformations to generate an executable simulation model from a UML graphic specification using the *Discrete Event Specification (DEVS)* formalism [4]. DEVS provides a system theoretic foundation to execute models using DEVS simulation protocol. DEVS models are inherently hierarchical in nature and consist of two classes of models i.e. atomic and coupled. Atomic models contain the state machine of a component and coupled models, which acts as a collection box composing the model structure, contains many atomic and coupled models leading to a hierarchical design.

Many different paradigms like System Entity Structure (SES) and DEVS hierarchical modeling can very well be used to interface with UML structure diagrams. However, the problem comes at the level of atomic components that contain finite state machines as their behavior model. Although the UML specification contains statecharts, their mapping with the DEVS state machine results in augmentation of UML statecharts with new added information for which there is no UML specification present, for example, timeouts for each state, a.k.a. *time advance* in DEVS. This problem has been highlighted by Mittal in [5] and an argument is presented that DEVS is more rigorous when it comes to modeling a state-based system. DEVS is more known in academic community while UML is widely practiced in the industry. The aim of this paper is to specify the graphic language so that systems engineering modelers may learn how to apply and use UML to build DEVS models, both structure and behavior. At the structure design level, we utilize UML component, package and class diagrams. At the behavior design level, we use UML use case, sequence, timing and state machine diagrams. In order to provide widespread adoption of the proposed executable UML, we coin the acronym ***eUDEVS*** that stands for executable UML based on DEVS. While the basic conceptual mapping is provided in [5], the present research provides a detailed implementation.

The proposed UML-based M&S method takes three steps. First, we synthesize the static structure defined by a DEVS or SES or UML model. Second, we specify its behavior using XML-Based *Finite Deterministic DEVS State Machine (XFD-DEVS SM)* model [6]. At this stage the models are totally platform independent. Lastly, we take these PIMs and autogenerate the Platform Specific Models (PSM) using various XML-based transformations that depend on the DEVS simulation engine syntactical requirements [7], [8], [9], [10]. The present research uses two simulation engines to illustrate the PIM-to-PSM transformations.

There are many *UML Computer Aided Software Engineering (UML CASE)* tools such as IBM Rational Rose, Poseidon, etc, and all of them provide simulation functionalities, tracing states change or signal invocation. All these tools have proprietary simulation engines. Further, most of the simulation engines are not extensible towards performance related requirements. For example, a typical OPNET model takes days to complete an execution. A parallel simulation engine would be needed but due to the proprietary nature, such extensions are not possible. For our purpose, we need an open-source specialized simulation engine which can take over the details of the simulation process (event management, simulation time management, etc.), provides extensibility and is an implementation of DEVS formalism. We are henceforth using DEVSJAVA version 3.0 [7] and Microsim/JAVA [11] to develop our case.

After demonstrating the transformation of UML models into DEVS component based system, we will go a step further to make these components fall under an overarching DEVS Unified Process (DUNIP) that is based on bifurcated model-continuity based lifecycle process [12]. DUNIP allows the development of test suite in parallel with the development of the system model. It is a logical extension which allows UML based models, once made executable form a component in a unified process much like the IBM Rational Unified Process.

To provide an overview, this research has the following objectives:

1. To unify the UML community with DEVS community
2. To facilitate the execution of UML models, especially behavior models, using DEVS formalism
3. To demonstrate that behavior can be represented using an XML-based DEVS formalism *with some caveats and limitations*

4. To illustrate that UML and DEVS can be cross-transformed
5. To make UML models as a component in an overarching systems engineering-based DEVS Unified Process.

The paper is organized as follows. In Section 2, we introduce the underlying technologies used, such as System Entity Structure (SES) [13], DEVS, UML, FD-DEVS and XFD-DEVS SM. It gives an overview of SES ontology that is used to develop metamodels of UML, DEVS and eUDEVS. Section 3 introduces the related work. In Section 4, we describe the eUDEVS metamodel and various mappings that allow UML to be transformed into DEVS. Section 5 demonstrates the application of this approach to the development of a DEVS model and the needed transformations using XML/XSLT technologies. Section 6 describes the DUNIP framework and integrates eUDEVS within it. Section 7 provides a detailed example for the entire process. Section 8 discusses the impact of the current research with respect to the related work and Section 9 provides conclusions and future work.

## 2. Background

Before we start giving an overview on UML and DEVS, we would like to elaborate on an ontology called System Entity Structure (SES). We will bridge UML and DEVS within this ontology framework. Then we shall discuss the Finite Deterministic DEVS and its XML implementation XFD-DEVS that is the key enabler for a platform independent state machine specification. This section will describe all the underlying technologies related to UML-DEVS executable model.

### 2.1 SES

System level design is made possible by System Entity Structure (SES) [13], [14]. The SES is a high level ontology framework targeted to modeling, simulation, systems design and engineering. Its expressive power, both in strength and limitation, derive from that domain of discourse. An SES is a formal structure governed by a small number of axioms that provide clarity and rigor to its models. The structure supports hierarchical and modular compositions allowing large complex structures to be built in stepwise fashion from smaller, simpler ones. Tools have been developed to transform SESs back and forth to XML allowing many operations to be specified in either SES directly or in its XML guise. The axioms and functionality based semantics of the SES promote pragmatic design and are easily understandable by data modelers. Together with the availability of appropriate tool support, it makes development of XML Schema transparent to the modeler [14]. Finally, SES structures are compact relative to equivalent Schema.

An SES is a labeled tree. Nodes of the tree are called *entities*, *aspects*, *specializations*, and *multiple decompositions*. An *entity* node represents a real world object (which can be independent or can be identified as a component of some decomposition or specialization of a real world object). *Aspects* are decomposition types, or in other words decomposition views. It is represented by a single bar (|). *Specialization* is a taxonomic relationship, i.e., a particular way of classifying an entity, which is represented by a double bar (||). It depicts 'types' or 'examples' any particular entity can take. *Multiple decomposition* is a special type of decomposition (aspect) that is used to represent entities whose number in a system may vary. It is used when similar kind of entities need be grouped together. It is represented by a triple bar (|||). The tautology will become clearer in a minute when we will show how UML metamodel can be depicted using SES.

When SES is applied to any specific design problem, a domain model is constructed that contains the entire solution set in all its permutations and combinations. To make is usable, in fact, executable, this SES needs to be 'pruned'. The pruning process is based on various design constraints, requirements and objectives as stated in the problem requirements document. A pruned SES is a design that can be realized in the real world as well as in an executable domain specific model. To construct a desired simulation model to meet the design objective, the pruning operation is used to reduce the SES to a pruned entity structure, PES [13]. The pruned entity structure can be transformed into a composition tree and eventually synthesized into a simulation model. We must point out that this developed simulation model doesn't contain the behavior specifications. Here a simulation model means that the SES representation for a given problem domain can be reduced to a platform specific implementation that can participate in various other extensible frameworks that can be made executable. Zeigler proposed the System Entity Structure (SES) [13], [14], and the SES is a theory to design systems hierarchically and structurally. The SES is a system entity that represents the real system enclosed within a certain choice of system boundary. The SES includes entities and their relationships. We will show in this current research how SES ontology can be used to specify a behavior model.

### 2.2 UML

In the field of software engineering, the Unified Modeling Language is a standardized visual specification language for object modeling [15]. UML is a general-purpose modeling language that includes a graphical notation used to create an abstract model of a system, referred to as a UML model.

UML is officially defined at the *Object Management Group (OMG)* [15] by the UML meta-model. The UML meta-model and UML models may be serialized in XML. UML was designed to specify, visualize, construct, and document software-intensive systems. However, UML is not restricted to modeling software. UML is also used for business process modeling, systems engineering modeling and representing organizational structures. UML has been a catalyst for the evolution of model-driven technologies. By establishing an industry consensus on a graphic notation to represent common concepts like classes, components, generalization, aggregation, and behaviors, UML has allowed software developers to concentrate more on design and architecture.

UML models may be automatically transformed to other representations (e.g. Java) by means of XSLT or QVT-like transformation languages, supported by the OMG. In addition, UML is extensible, offering the following mechanisms for customization: profiles and stereotype. The semantics of extension by profiles have been improved with the UML 2.0 major revision.

UML 2.0 has 13 types of diagrams, which can be categorized hierarchically as shows Figure 1. Structure diagrams emphasize what things must be in the system being modeled and include class diagram, component diagram, composite structure diagram, deployment diagram, object diagram and package diagram. Behavior diagrams emphasize what must happen in the system being modeled. They include activity diagram, state machine diagram and use case diagram. UML also includes interaction diagrams, a subset of behavior diagrams, used to define the flow of control and data among the entities in the system being modeled. Interaction diagrams are communication diagram, interaction overview diagram, sequence diagram and timing diagram.

The UML metamodel in Figure 1 is put together within the SES ontology. Single bars (|) implies that the entity is decomposed into the following entities. For example, UML Diagram is made of two different perspectives, i.e. Structure and Behavior diagrams. For an effective UML model, both the diagrams are necessary. Similarly, Behavior diagram is made of Activity Diagram, Interaction diagram, Use Case diagram and State Machine diagram. However, there can be many ways in which Interaction diagrams can be implemented. Depending on the project requirements, an Interaction diagram 'can be' of type Communication diagram, Sequence diagram, Interaction Overview diagram and Timing diagram. The choice is denoted by a double bar (||) in SES ontology. Likewise, the Structure Diagram consists of Class diagram, Component diagram and Deployment diagram. In addition to this decomposition perspective of Structure Diagram, it is specialized as a composite Structure diagram that is made up many entities of type Component diagram. A Component diagram is made of Package diagram that contains many Class diagram entities.

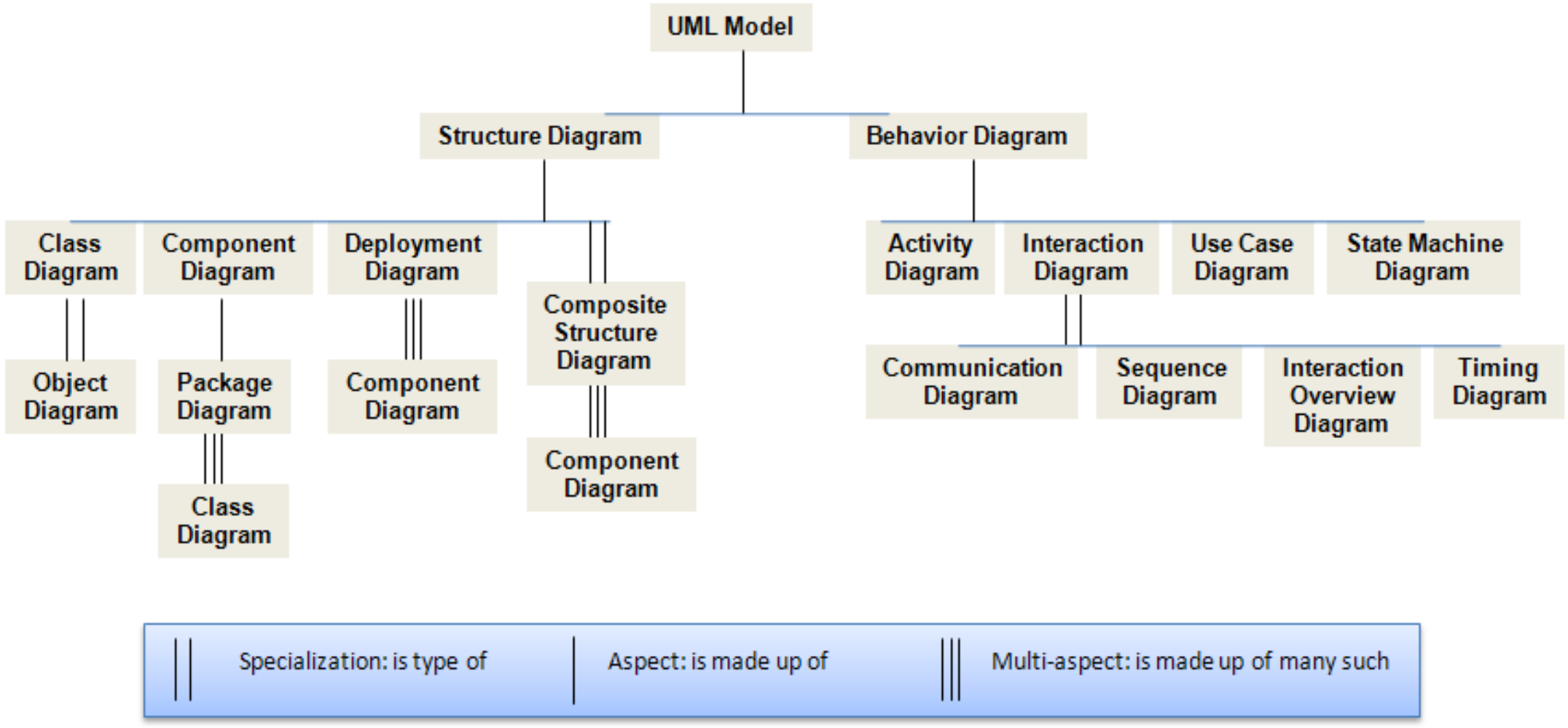

**Figure 1.** UML Metamodel

### 2.3 DEVS

The *Framework for M&S* as described in [4], establishes *entities* and their *relationships* that are central to the M&S enterprise (see Figure 2). The entities of the framework are *source system, experimental frame, model,* and *simulator;* they are linked by the *modeling* and the *simulation* relationships. Each entity is formally characterized as a system at an appropriate level of specification within a generic dynamic system. See [4] for detailed discussion.

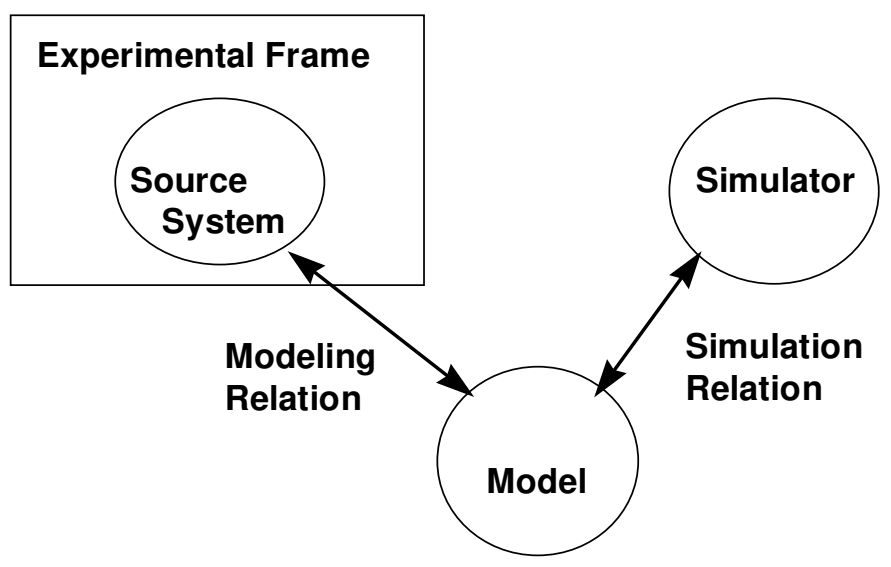

**Figure 2.** Framework Entities and Relationships

DEVS formalism consists of this framework as depicted in Figure 2 above. The *experimental frame* consists of various requirements, constraints and design decisions that are put on the source system for realizing the *model*. We will ignore the modeling relation for the time being. In this present research we will focus on the *model* and the *simulator* aspect only. The simulation relation is implemented as DEVS simulation protocol. For more details on the simulation protocol, please refer [4] as it again is not the focus of this research. We are primarily concerned with the *model* aspect in Figure 2 and how UML can provide us the means to construct models and vice versa.

DEVS Models are of two types: atomic and coupled. The atomic model is the irreducible model definition that specifies the behavior for any modeled entity. The coupled model is the aggregation/composition of two or more atomic and coupled models connected by explicit couplings. The formal definition of parallel DEVS (P-DEVS) is given in [4]. An atomic model is defined by the following equation:

$$M = \langle IP, OP, X, S, Y, \delta_{int}, \delta_{ext}, \delta_{con}, \lambda \rangle$$

where,

- $IP, OP$ are the set of input and output ports
- $X$ is the set of input values
- $S$ is the state space
- $Y$ is the set of output values
- $\delta_{int} : S \rightarrow S$ is the internal transition function
- $\delta_{ext} : Q \times X^b \rightarrow S$ is the external transition function
  - $Q = \{(s,e) : s \in S, 0 \le e \le ta(s)\}$ is the total state set, where $e$ is the time elapsed since last transition
  - $X^b$ is a set of bags over elements in $X$
- $\delta_{con} : S \times X^b \rightarrow S$ is the confluent transition function, subject to $\delta_{con}(s, \varnothing) = \delta_{int}(s)$
- $\lambda : S \rightarrow Y$ is the output function
- $ta(s) : S \rightarrow \Re_0^+ \cup \infty$ is the time advance function.

Two state variables are usually present in the state space of an atomic model: “phase” and “sigma”. In the absence of external events the system stays in the current “phase” for the time given by “sigma”.

The formal definition of a coupled model is described as:

$$N = \langle IP, OP, X, Y, D, EIC, EOC, IC \rangle$$

where,

- $IP, OP$ are the set of external (not coupled) input and output ports
- $X$ is the set of external input events
- $Y$ is the set of output events
- $D$ is a set of DEVS component models
- $EIC$ is the external input coupling relation
- $EOC$ is the external output coupling relation
- $IC$ is the internal coupling relation.

The coupled model $N$ can itself be a part of component in a larger coupled model system giving rise to a hierarchical DEVS model construction.

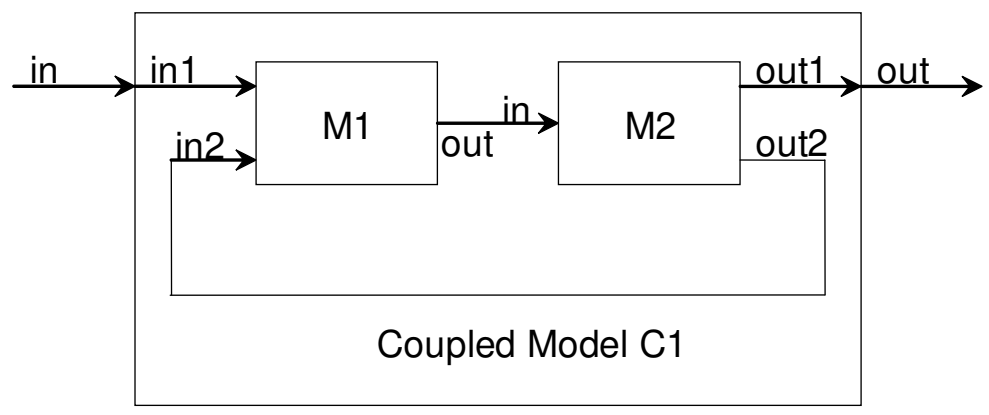

**Figure 3.** Coupled DEVS model

Figure 3 shows a coupled DEVS model. M1 and M2 are DEVS models. M1 has two input ports: “in1” and “in2”, and one output port: “out”. The M2 has one input port: “in1”, and two output ports: “out1” and “out2”. They are connected by input and output ports internally (this is the set of internal couplings, IC). M1 is connected by external input “in” of Coupled Model to “in1” port, which is an external input coupling (EIC). Finally, M2 is connected to output port “out” of Coupled Model C1, which is an external output coupling (EOC). Similarly, this coupled model C1 with ‘in’ and ‘out’ ports can become a component in another coupled model

DEVS formalism has many object-oriented implementations such as in C++, Java, and C#. DEVS simulation engine has been implemented on many distributed frameworks such as CORBA, P2P, Grid, High Level Architecture (HLA), Remote Method Invocation (RMI) [16], [17], [18], [19] and most recently Service Oriented Architecture (SOA) [20], [12]. The DEVSJAVA and Microsim/JAVA are Java based DEVS modeling and simulation environment. They provide the advantages of Object Oriented framework such as encapsulation, inheritance, polymorphism, etc. The Java implementation of DEVS manages the simulation time, coordinates event schedules, and provides a library for simulation, a graphical user interface to view the results, and other utilities. Detailed descriptions about DEVS Simulator, Experimental Frame and of both atomic and coupled models can be found at [4] and [11].

### 2.4 XML-Based Finite Deterministic DEVS

As the name implies, this kind of DEVS employs the finite and deterministic properties of the constituent elements. The sets of 'events' and 'states' are finite and all the characteristic functions associated are deterministic [21]. It is a 7-tuple formalism specified as below,

M=<incomingMsgSet, outgoingMsgSet, stateSet, timeAdvance, internalTransition, externalTransition, output>

where,

- incomingMessageSet, outgoingMessageSet, StateSet are finite sets
- timeAdvance:
  StateSet → R0,∞+ (the positive real with zero and infinity)
- internalTransition: StateSet → StateSet
- externalTransition:
  StateSet × incomingMessageSet → StateSet
- output:
  StateSet → 2outgoingMsgSet (the set of subsets of outgoingMsgSet)

Mittal has recently proposed a novel way to automate the DEVS state machine specification process [12]. In this approach, any state machine can be looked upon as the superposition of two behaviors. The first cycle is the default execution of the machine, wherein it receives no external inputs. These are the "internal" transitions of the component. The second behavior, which can spawn multiple cycles stems from the actions resulting from reception of various inputs in various states. These are the induced "external" transitions of the component. DEVS categorically separates these two behaviors in its formal $\delta_{int}$ and $\delta_{ext}$ specification respectively. Mittal also proposed a W3C XML Schema [6] based on FD-DEVS [21].

DEVS formalism has been extensively researched from the viewpoint of simulation and execution for over 30 years. One of the major problems on verification of DEVS state space has been accomplished only recently by abstracting the infinite state behavior of DEVS as a finite reachable graph [22], [23]. This verifiable model is called FD-DEVS [21] which is a subset of DEVS formalism. According to latest developments XFD-DEVS has been implemented as an XML representation giving it a platform independent structure by Mittal in his doctoral thesis [12]. This XML-based approach as verified by Moon's FD-DEVS [21] is XFD-DEVS. The XML based Atomic and coupled models are validated by Atomic and coupled schemas that were also developed towards DEVS Standardization processes (see Appendix A and C).

The formal specification of FD-DEVS employs the prescribed semantics of DEVS itself, as a specification of subclass within the broader class of I/O Dynamic systems [21]. In addition, XFD-DEVS has been extended towards a natural language processing (NLP) interface where in the state machine can be specified by simple English statements [6].

We've developed a tool called XFD-DEVS workbench that provides a template based state requirement definition process. This process is executed as follows. It starts with user specifying the name of the atomic model. On clicking the 'Finite State Time-Advance' button as in Figure 4, State space is specified wherein the timeout for each state is provided by the user affront. Next, the user specifies the two behavior cycles viz., the default internal behavior and the externally induced behavior, as shown in Figure 5 and 6 respectively. The user is required to specify the state machine in a tabular format which is transformed into a Platform Independent Model (PIM) in XML (Figure 7). It is in Figure 5 and 6 that the behavior is separated into the default 'internal' behavior and the 'external' induced behavior for the atomic component under design. Finally, on clicking the 'Generate FD-DEVS' button (Figure 7) the tabular data is realized as an XML-based PIM conforming to Atomic and Coupled schemas for XFD-DEVS (See Appendix A and C). This XFD-DEVS PIM is readily realized in platform specific model (PSM) implementation in DEVSJAVA (Figure 8), DEVS.net and Microsim/JAVA.

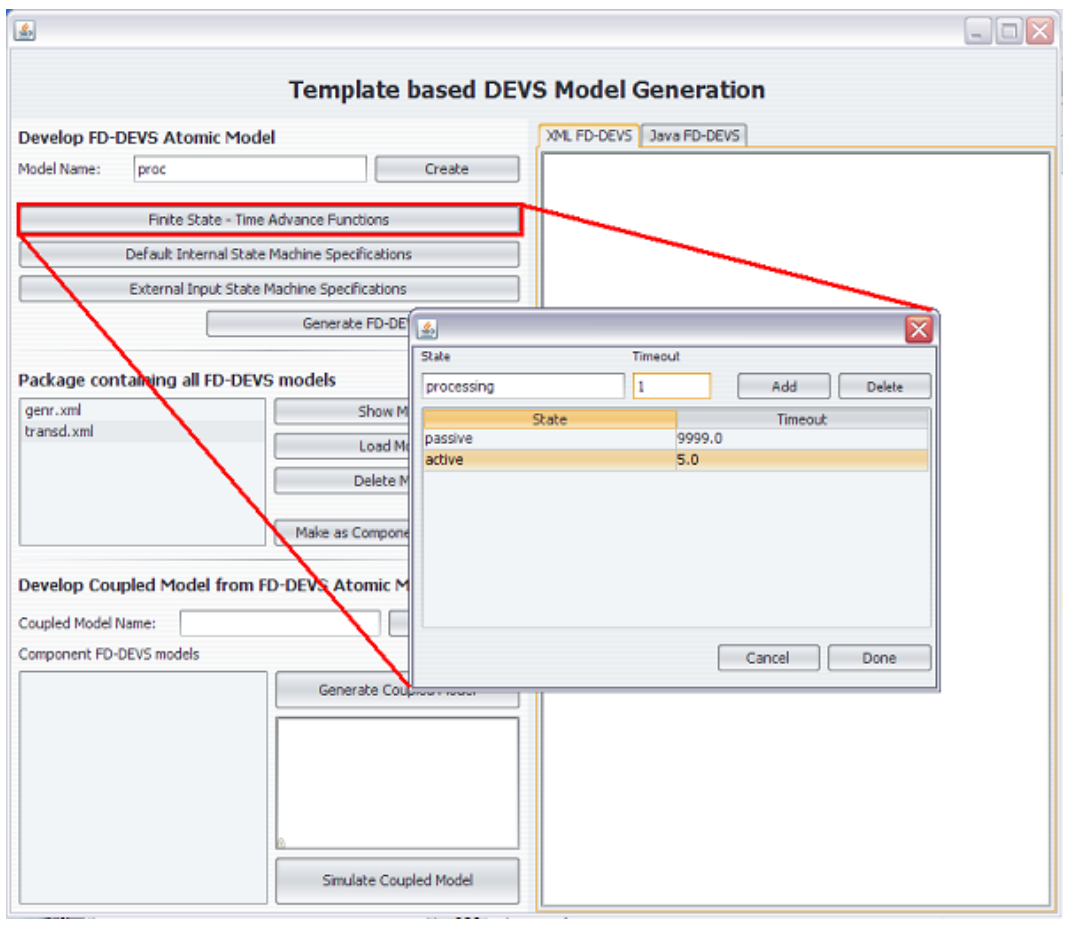

**Figure 4.** XFD-DEVS state space with each state having a finite timeout. Infinity is allowed

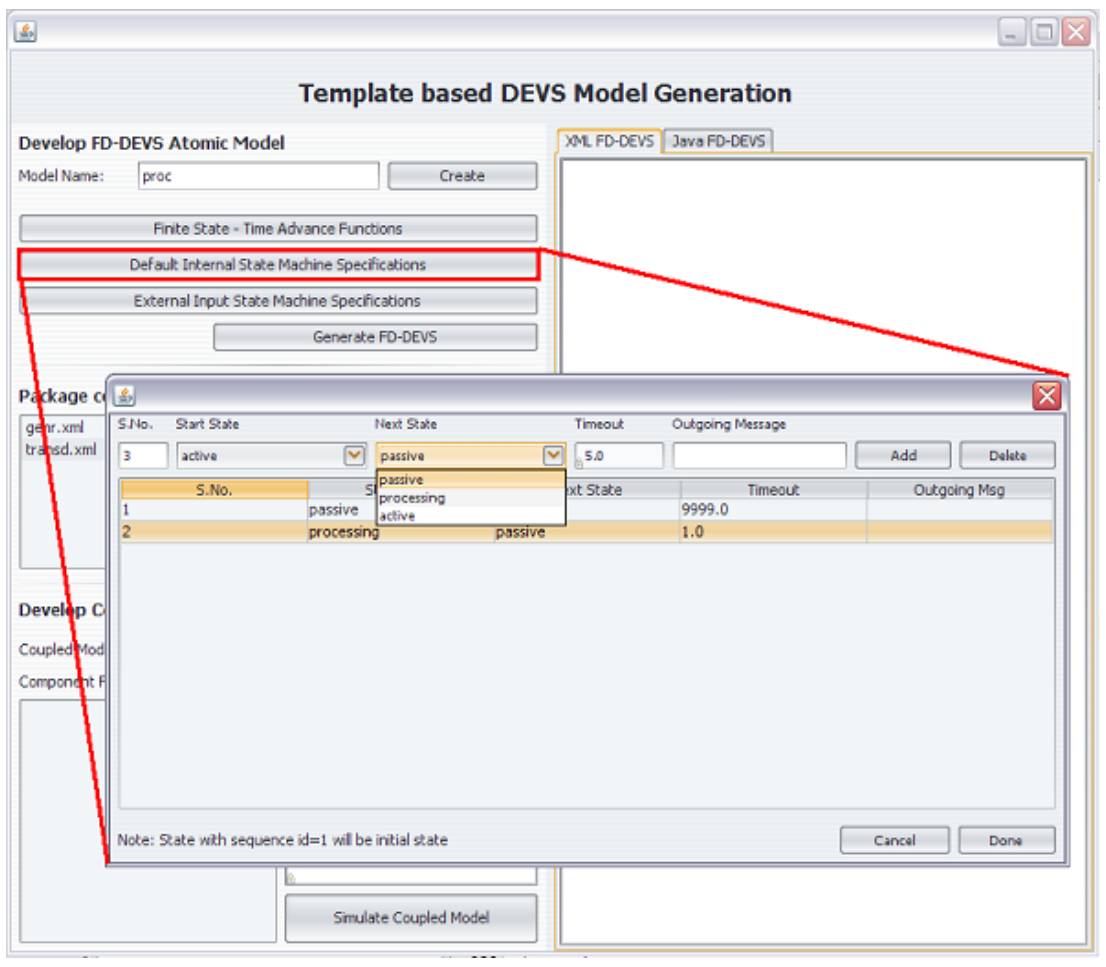

**Figure 5**. Internal behavior table

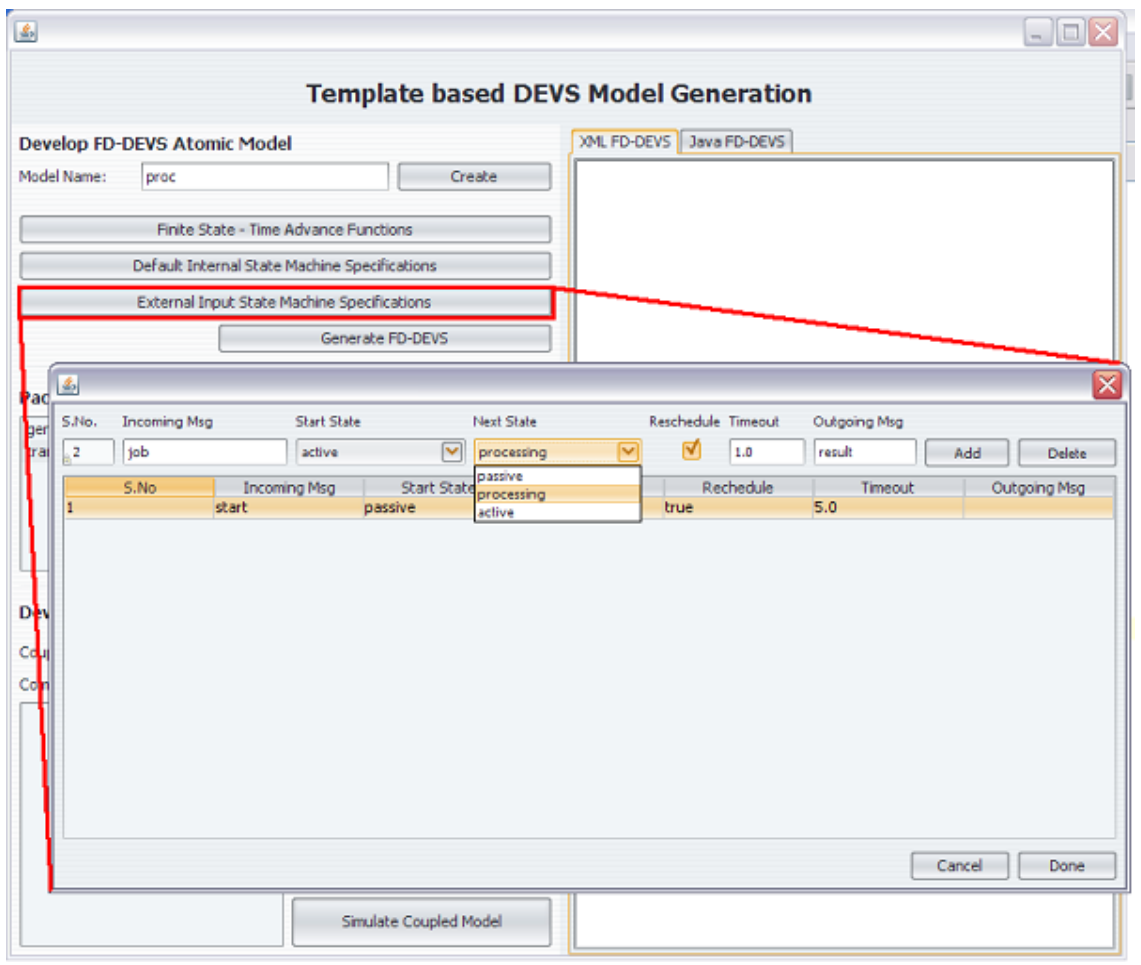

**Figure 6**. External Behavior Table

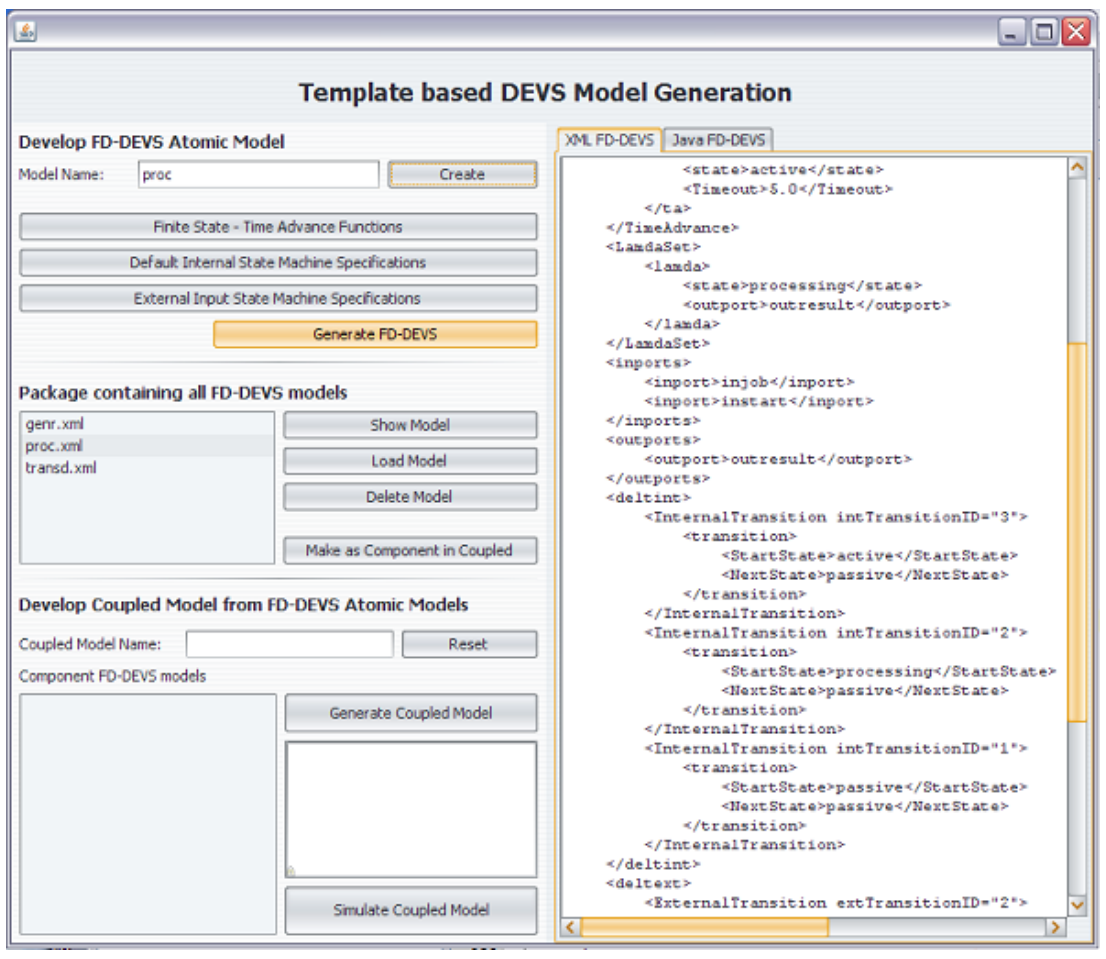

**Figure 7.** PIM XFD-DEVS generated model

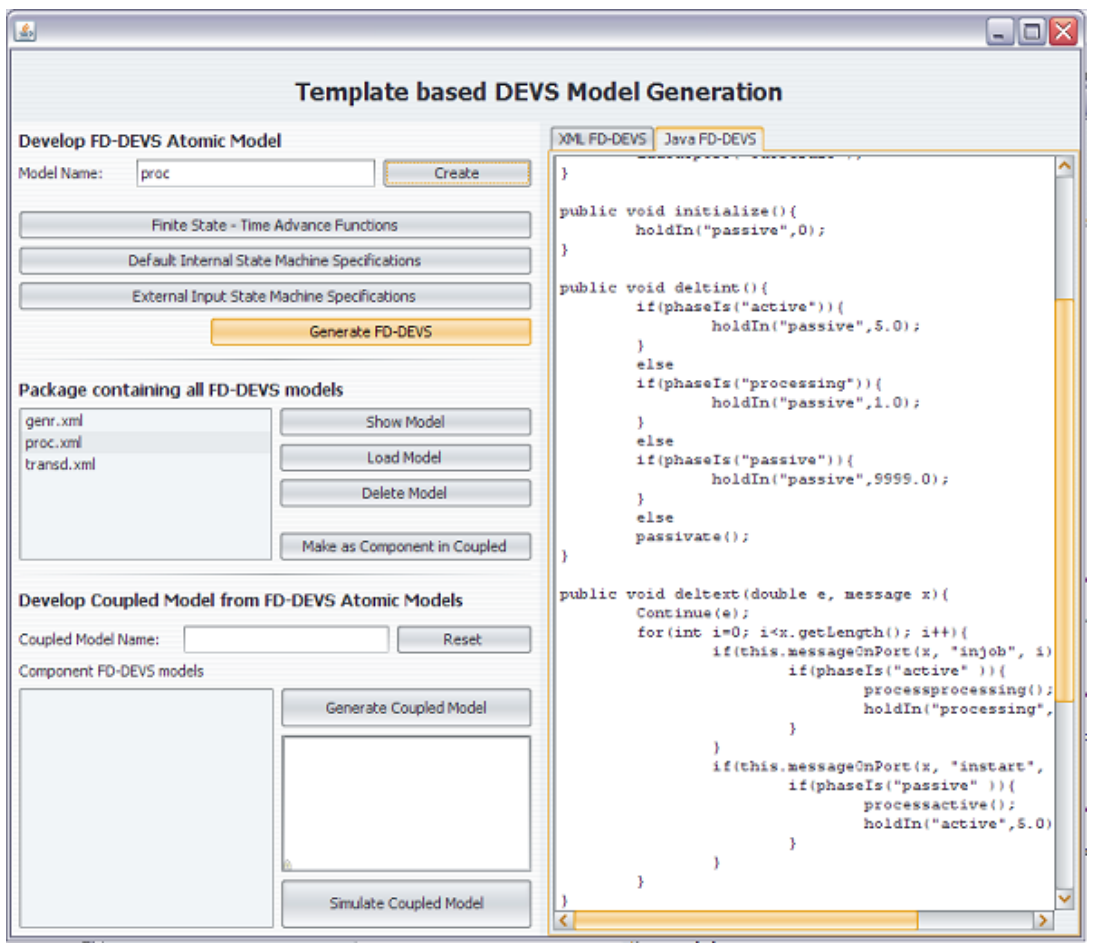

**Figure 8.** PSM in Java

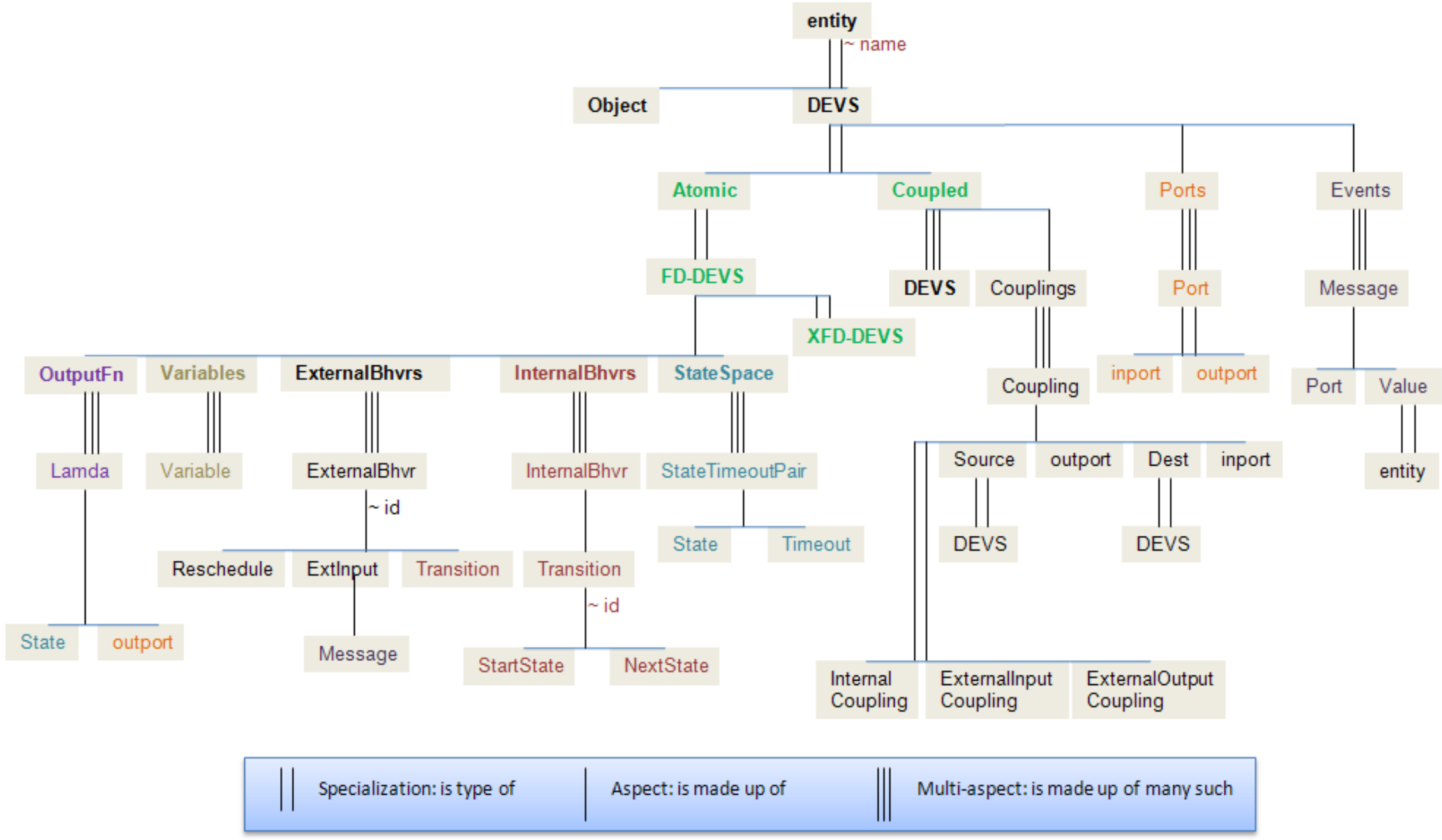

**Figure 9.** DEVS Metamodel

### 2.5 DEVS Metamodel

The DEVS metamodel as per SES ontology is depicted in Figure 9 above. It starts with an *entity* that can be any *Object* or a DEVS *model*. DEVS is made of *models*, *Ports* and *Events*. *Coupled* is recursively made of DEVS *models* and *Couplings*. XFD-DEVS is a type of Atomic model and is made of *OutputFn*, *Variables*, *ExternalBehaviors*, *InternalBehaviors* and *StateSpace*. *Couplings* is a multi-aspect entity shown by triple bar (|||) and is made of many such *Coupling* entities. A *Coupling* is made of *Source*, *Destination*, *outport* and *inport* objects. A *Coupling* can be *Internal coupling*, *ExternalInput coupling* or *ExternalOutput coupling* based on DEVS formalism [4]. *Source* and *Dest* can be DEVS models. *Port* can be *inport* or *outport*. *Events* are made of many such *Message* entities. A *Message* is made of *port* and *value*. A *Value* can be an *entity*.

## 3. Related work

In our earlier effort [10], we proposed a W3C XML Schema for DEVS coupled models that enable the user to provide the system component structure in hierarchical DEVS notation. This describes the structure and a limited behavior of any DEVS model. However, it is not a graphic solution, because the model must be written in XML to be transformed to the simulation engine.

There are other approaches to represent DEVS models, such as the Scalable Entity Structure Modeler with Complexity Measures [24]. It is suitable for developing component-based hierarchical models. It offers a basis for modeling behavioral aspect of atomic models by providing the structural specification and storage of the model using XML but, this approach is very close to the simulation expert instead of the domain expert. Further, there are no means to develop the atomic state machine or behavioral model explicitly. This is largely a structure tool and needs further work to represent atomic models using XML.

In the UML-based M&S domain, several authors have approached this subject from various perspectives. Choi [25] utilizes UML sequence diagrams to define the system behavior. In [26], eight steps are introduced in order to make DEVS models using UML, but in these cases they need many human decisions to transform the model. A formal mapping from DEVS to UML is presented by Zinoviev in [27]. Within this mapping input and output ports are mapped to UML events. DEVS state variables that are non-continuous are mapped to UML states, and DEVS state variables that are continuous are mapped to attributes of UML states. Zinoviev also employs a combination of a special timeout event and use *after* events for handling internal transitions. The mapping presented is elegant. However, his UML representation is not intended to provide a unified representation on top of a modeling formalism for the purpose of composition but as a replacement for the original DEVS specification. In [28], Huang and Sarjoughian present a mapping for coupled models into UML-RT structure diagrams, but the use of the UML Profile for Schedulability, Performance and Time Specification (OMG 2005) is unnecessary when mapping from DEVS to UML. They conclude that UML-RT is not suitable for a simulation environment, and they assert that the design of software and simulation is inherently distinct as we address in this paper. In [29], Borland and Vangheluwe develop a methodology to transform hierarchical state-charts into DEVS. In [30], the formal transformation of timed input/output automata into a DEVS simulation model is provided. However, the timed automata is hard to communicate and develop the simulation model. We recently proposed a UML-based M&S method, making use of UML state machine diagrams to define the system behavior, where by using the *State Chart eXtensible Markup Language (SCXML)* [31], time events are clearly defined in order to generate executable atomic models [32]. Finally, an informal mapping from DEVS to an equivalent STATETMATE Statechart is presented in [33]. Schulz et al note that DEVS has greater expressive capabilities than state charts [34] and that any DEVS model can be represented via STATEMATE Activity Charts along with an appropriate naming convention for events.

Extending the domain to the Model Driven Architecture (MDA) paradigm, Tolk and Muguira [35] show how the complementary ideas and methods of the High Level Architecture (HLA) and DEVS can be merged into a well-defined M&S application domain within the MDA framework. HLA is distributed simulation architecture regardless of computing platforms. It provides an interface that each constituting simulation engine must conform to participate in a distributed simulation exercise. While it is widely used in the defense industry, its adoption into the industry has been prohibited by its lack of expressive power.

As it is clear, none of this work has attempted to provide a unifying framework to bridge the UML-DEVS gap and executable UML in particular. In this paper we analyze not only the specification of both the structure and behavior of DEVS models in UML, but also a general-purpose modeling procedure for DEVS models, supporting the specification, analysis, design, and verification and validation of a broad range of DEVS-based systems.

## 4. Mapping UML to DEVS in eUDEVS

### 4.1 Overview

Modeling is an art of abstraction of real systems to generate the behavior by specifying a set of instructions, rules and equations to represent the structure and the behavior of the system. The structural elements of a system include the components of the system and their interactions (inputs, outputs, connections, etc). The behavioral elements include the sequence of interactions, the timing constraints of the interactions, and the operations of each component. Modeling provides the means of specifying the structure of a system, behavior of a system over time, and the mechanism for executing the instructions, rules, or equations.

We provide a new approach for both the structural and behavioral graphic description. Currently it is common practice for systems engineers to use a wide range of modeling languages, tools and techniques on large systems projects. Since UML unified the modeling languages used in the software industry, our approach is intended to unify the diverse modeling languages currently used by DEVS-based systems engineers. It will improve communication among the various stakeholders who participate in the systems development process and promote interoperability among DEVS-based M&S tools.

Our DEVS UML diagram taxonomy is shown in Figure 10. This diagram has its origin in figure 1 but focuses on only those UML elements that contribute towards a DEVS-based system model. A DEVS model is defined by means of structure and behavior diagrams. To define the structure we make use of UML component diagrams, package diagrams and/or class diagrams. To define the behavior we utilize UML use case diagrams, state machine diagrams, sequence diagrams and/or timing diagrams. In order to develop an interface between DEVS and any other modeling framework, such as UML here, we should first enumerate the information that is needed to develop a DEVS model.

1. Entities as Objects and their hierarchical organization
2. Finite State Machines (FSMs) of atomic models
3. Timeouts for each of the phases (States) in atomic models.
4. Entity interfaces as input and output ports
5. External incoming messages at Entity's interface at specified duration in specific State
6. External outgoing messages at Entity's interface at specified duration in specific State
7. Coupling information derived from hierarchical organization and interface specifications
8. Experimental Frame specifications

Having known the information needed to develop DEVS, the following sub-sections will show how this information is extracted from UML elements. The Experimental Frame specifications mapping is outside the scope of this paper and more information about their design can be seen at [20].

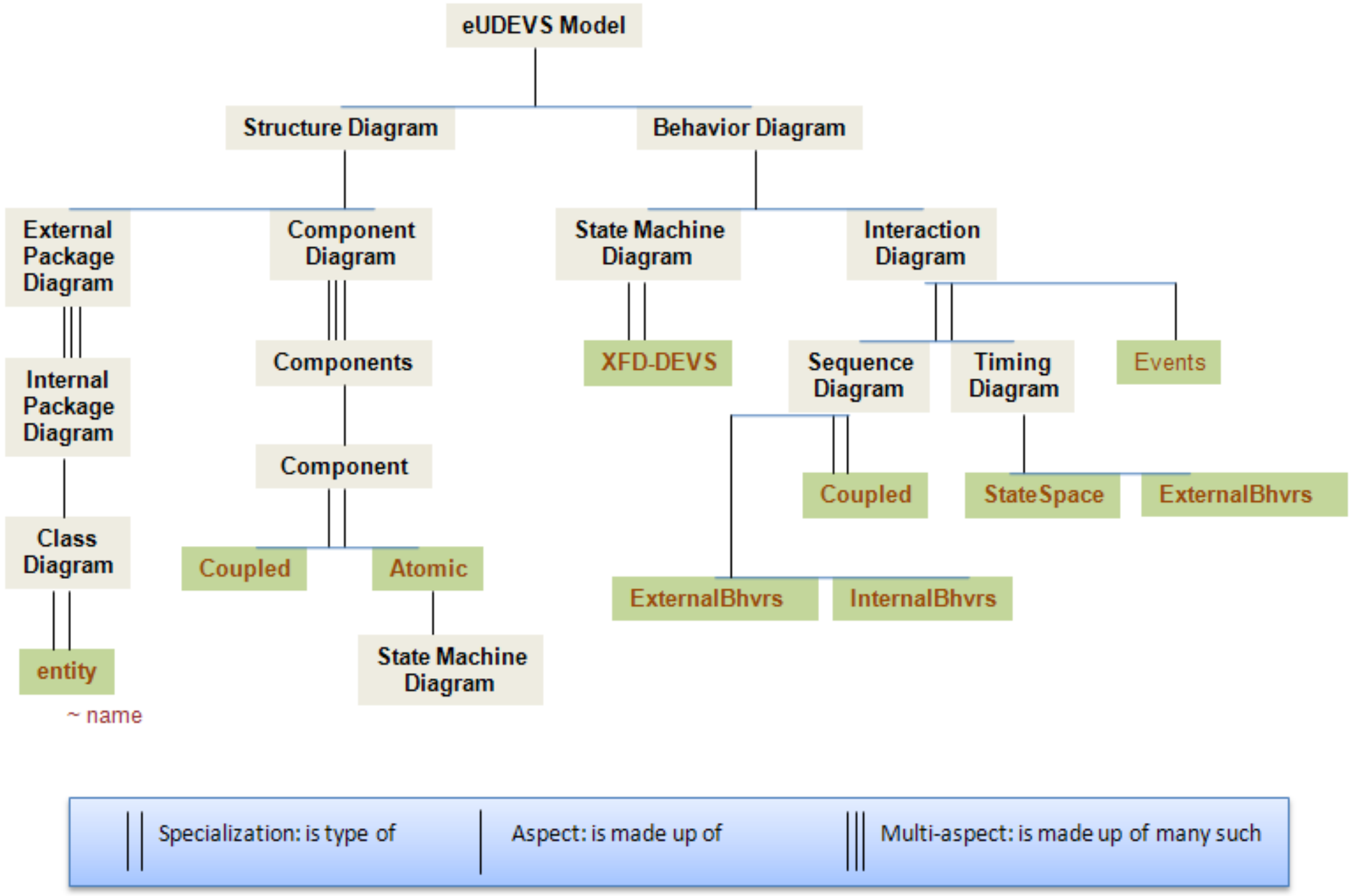

**Figure 10**. Executable UML using DEVS: eUDEVS Metamodel

### 4.2 DEVS UML Structure in eUDEVS

This section defines the static and structural constructs used to describe DEVS UML structure diagrams, including the component diagram, package diagram and class diagram.

*4.2.1 Component diagrams*

UML component diagrams have evolved substantially from version 1.x of the language to the current 2.0. In the latest version of the language, deployment, object, and component diagrams have been merged into a single class of deployment-object-component diagrams. These new diagrams have a rich language suitable for elaborated models, but for the purpose of this research we need to mention only components, delegation connectors, interfaces, and ports.

A UML component diagram is a set of components, ports, connections and interfaces. UML components may contain sub-components in a hierarchical fashion similar to DEVS coupled models containing models. Each UML component has a set of externally visible ports. UML components may be connected to one another and attached via ports similar to DEVS. UML ports can be unidirectional (input or output) or bidirectional. The direction of the port is defined by the type(s) of its interfaces. Required interfaces ("antennas") define output ports, and provided interfaces ("lollipops") define input ports. In DEVS connections between ports are unidirectional. Therefore, in a UML component diagram, all ports should be unidirectional, i.e., a port may either provide an interface or require an interface but not both since this implies bi-directionality. Components may be connected in two different ways. First we can connect provided and required ports by a so called assembly connector. In contrast, the delegation connector connects two ports of the same type between a component and one of its sub components. Thereby, hierarchical components become possible.

Using *parallel* DEVS as the target formalism the mapping of component connections to model couplings and their association with model ports becomes relatively easy. Each component connection maps to a set of couplings $c \in \{EIC \cup EOC \cup IC\}$. Assembly connectors map to internal couplings and delegation connectors map to external input and output couplings.

In UML, ports may have a multiplicity greater than one; this is not the case in DEVS (and hence not allowed in our UML component diagrams). In UML, port may be unnamed; they must be named in DEVS and thus in our UML component diagrams. In UML, connectors need not attach to components (more correctly *parts*) via ports; this is not an option in DEVS and hence not an option in UML. If ports are specified to provide or require an interface, there should be only one such interface specified in the component diagram.

Finally, a UML component represents either another UML component diagram, or a UML state machine diagram. We will consider that a DEVS coupled model may only include components and a DEVS atomic model may only contain state machine diagrams.

A simple UML component diagram is shown in Figure 11, which corresponds to the DEVS coupled model shown in Figure 3. Figure 11 depicts two components M1 and M2. These components are in turn subcomponents of component M. Component M has one output port "out" with interface "Event1" and one input port "in" with the same interface. M1 has two input ports: "in1" and "in2" with interface "Event1", and one output port "out" with interface "Event2". The M2 has one input port: "in1" with interface "Event2", and two output ports: "out1" and "out2" with interface "Event1". Ports connected must send/receive compatible interfaces.

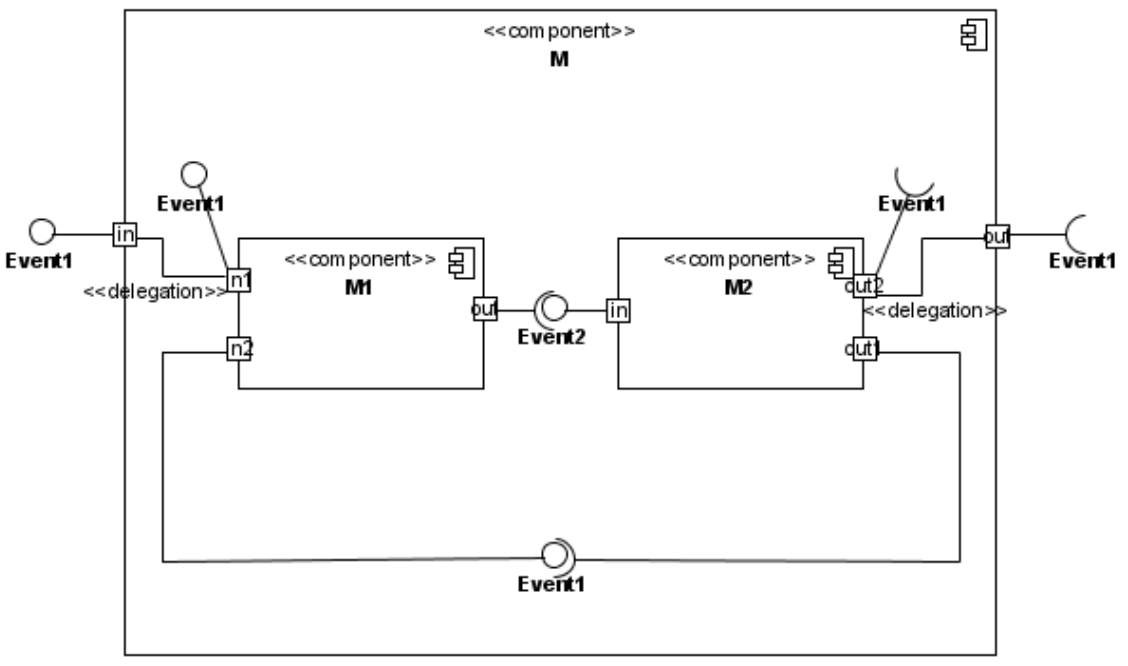

**Figure 11.** A DEVS UML Component Diagram

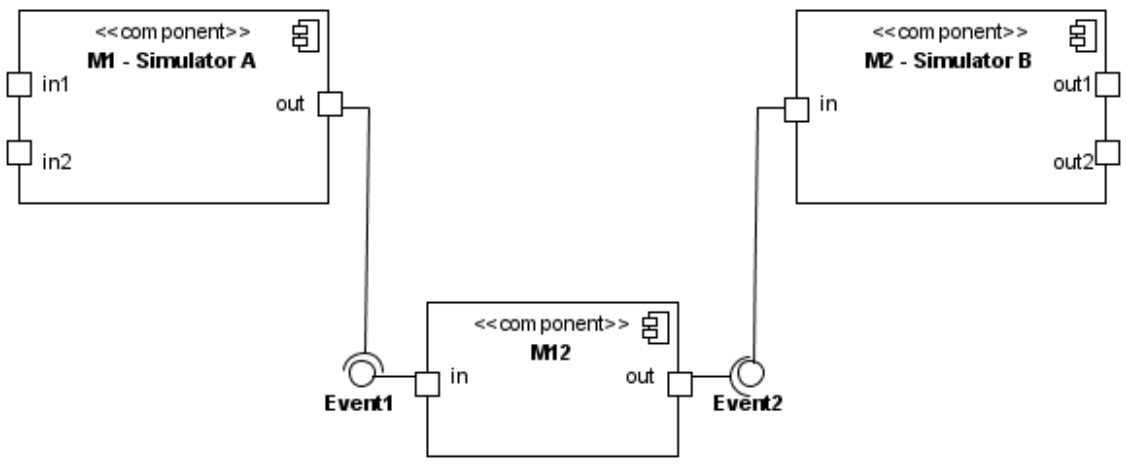

**Figure 12.** Interoperability between components

Interoperability between different simulation platforms is possible at this level of message abstraction using XML [36]. The message structure with a port-value pair is very well the foundation of interoperable and cross-platform system integration. The latest development of DEVS/SOA platform is an evidence of such XML based message passing between different DEVS components [20]. In Figure 11, M1 could be implemented using one simulation engine and M2 by another one. The only constraint is that messages (or events) sent by M1::out must satisfy the same interface that those received by M2::in. If this constraint is not satisfied, another component M12 should be defined allowing transformations to make interfaces compatible as shown in Figure 12. Table I shows the relation between DEVS structural formalism and UML component diagrams.

**Table I.** Relation between DEVS structural formalism and UML component diagram

| DEVS | UML Component diagram |
|---|---|
| **Atomic model (Component containing UML state machine)** | |
| *IP* | Ports with provided interface |
| *OP* | Ports with required interface |
| **Coupled model (Component containing components)** | |
| *IP* | Ports with provided interface |
| *OP* | Ports with required interface |
| *D* | Components |
| *EIC* | Input delegation connectors |
| *EOC* | Output delegation connectors |
| *IC* | Interfaces connected |

*4.2.2 Package diagrams*

In UML, a package diagram depicts how a system is split up into logical groups by showing the dependencies between them. As a package is typically thought of as a directory, package diagrams provide a logical hierarchical decomposition of a system. Packages are usually organized to maximize internal coherence within each package and to minimize external association among packages. With these guidelines in place, the packages are good management elements. Each package can be assigned to an individual or team, and the dependencies among them indicate the required development order.

To define the structure of an M&S-based system, we define two kinds of representations in terms of package diagrams: external and internal. Packages in external representation will contain only packages, i.e., the internal representation. Packages in internal representation will contain various class diagrams or models. The external representation encapsulates three parts of an M&S-based system: (1) simulation engines required to implement the system, (2) components utilized in the components diagram, and (3) supporting classes which are not executed directly by the simulation engine, such as port interfaces, extra data types, etc. Internal representation defines the external representation structure in more detail. At this layer, we require one package per component in the components diagram. To this end, we have defined several stereotypes: *engine*, *model*, *coupled*, *atomic* and *support* that correspond to designing packages for simulation engines, the entire DEVS model, coupled and atomic components, and supporting classes.

Figure 13 shows the external (left side) and internal representation (right side) of the component diagram in Figure 11. There are three packages, M, M1 and M2 in the internal representation which correspond to each component in the model. DEVSJAVA is the selected simulation engine to implement the model, and we need some port interfaces as supporting classes.

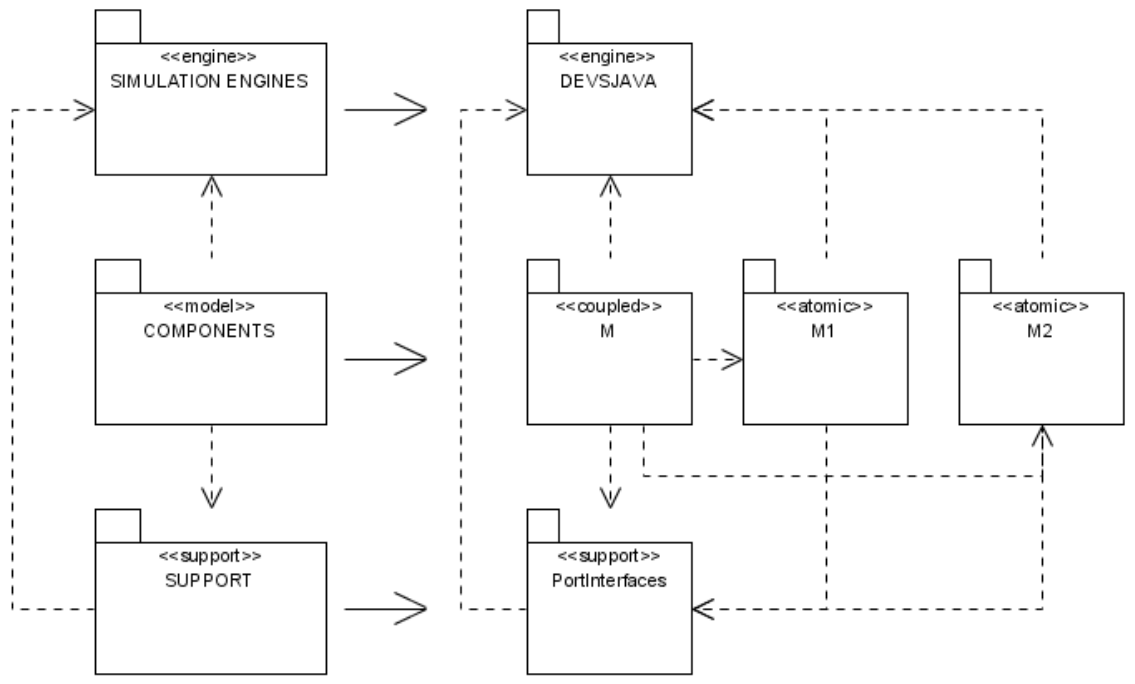

**Figure 13.** External and internal structure representation in package diagrams

In contrast to component diagrams, package diagrams do not define the structure of a DEVS system. They need the support of class diagrams and sequence diagrams to define both the structure and behavior.

*4.2.3 Class diagrams*

UML 2 class diagrams are the mainstay of object-oriented analysis and design. UML 2 class diagrams show the classes of the system, their interrelationships: inheritance, aggregation, and association, and the operations and attributes of the classes. Class diagrams are used for a wide variety of purposes, including both conceptual/domain modeling and detailed design modeling.

By using the information contained in the component diagram and/or package diagrams, we are able to distinguish classes that are directly used by the simulation engine from supporting classes that are not.

Figure 14 depicts an illustrative example of two object diagrams. Figure 14a shows the *M* package (as per the package design in last subsection, the package name is same as the component name). The class *CoupledM* represents a DEVS coupled model. *CoupledM* contains two atomic models: *AtomicM1* and *AtomicM2*, both coming from component packages *M1* and *M2*, respectively. Finally, since we are using the DEVSJAVA simulation engine, *CoupledM* extends the functionality of *Digraph*, which is the base class for coupled models in DEVSJAVA. Figure14b shows the *M1* package. The *AtomicM1* class is an atomic model, so it inherits *Atomic*, which is the base class used in DEVSJAVA to implement atomic models. *AtomicM1* sends and receives messages composed by elements that implement the interfaces *Event1* and *Event2*. Both interfaces must extend the *Entity* DEVSJAVA class as shows Figure 14b, since it is the base class used to send messages between components. Please note that for example, the DEVS component M which corresponds to the M package could be formed by more than one class. The example provided is quite simple and, in this case, every DEVS component is implemented using only one class.

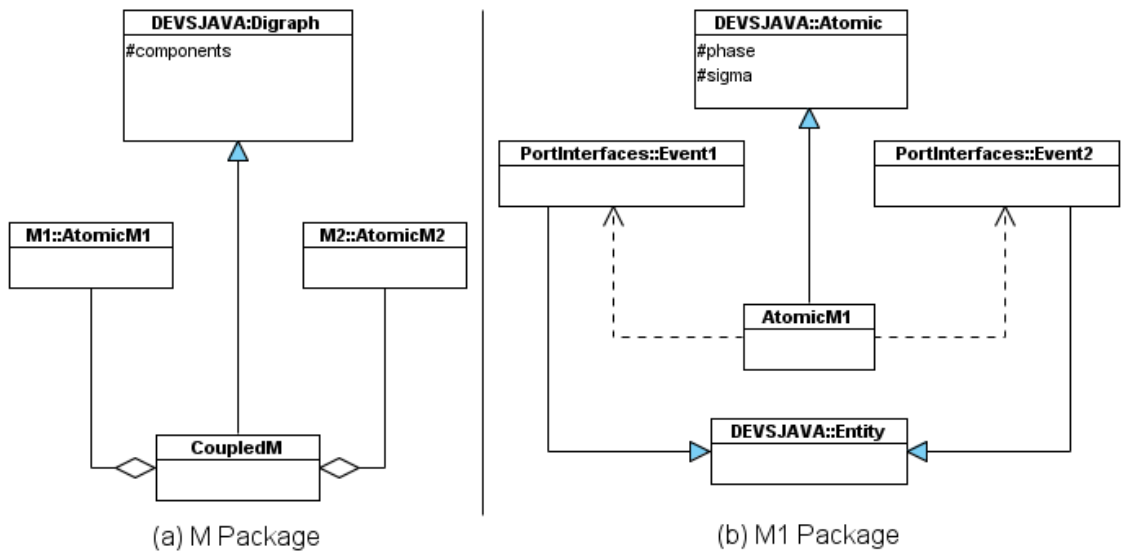

**Figure 14.** Two simplified class diagrams of the system

We have defined three diagrams which allow defining the structure of a DEVS-based system. Component diagrams are able to describe a DEVS model structure by themselves. However, package and class diagrams provide a software development point of view, which is important to design the modeling environment. Combining the information of these diagrams, we may generate the skeleton of such models, in terms of classes, attributes, empty member functions, etc. This generation of software artifacts is platform dependent though. Again, we must stress that at this stage we only have an executable structure model that is devoid of any behavior. The next section will address this issue.

### 4.3 DEVS UML Behavior

This section specifies the behavioral definition of a DEVS model in terms of use case diagrams, sequence diagrams, timing diagrams and state machine diagrams.

#### *4.3.1 Use case diagrams*

*Use case* diagrams describe behavior in terms of the high level functionality and usage of a system, by the stakeholders, and other members such as developers who build the system. The *use case* diagram describes the usage of a system (subject) by its actors (environment) to achieve a goal, which is realized by the subject providing a set of services to selected actors. The *use case* can also be viewed as functionality and/or capabilities that are accomplished through the interaction between the subject and its actors. *Use case* diagrams include the *use case* and actors and the associated communication between them. Actors represent classifier roles that are external to the system that may correspond to users, systems, and or other environmental entities. They may interact either directly or indirectly with the system.

The *use case* relationships are “communication,” “include,” “extend,” and “generalization.” Actors are connected to *use cases* via communication paths, which are represented by an association relationship. The “include” relationship provides a mechanism for factoring out common functionality that is shared among multiple *use cases*, and is always performed as part of the base *use case*. The “extend” relationship provides optional functionality, which extends the base *use case* at defined extension points under specified conditions. The “generalization” relationship provides a mechanism to specify variants of the base *use case*.

The *use cases* are often organized into packages with the corresponding dependencies between them.

Figure 15 depicts how *use cases* help delineate specific kind of goals associated with driving and parking a vehicle. In Figure 15 the “extends” relationship specifies that the behavior of a *use case* may be extended by the behavior of another (usually supplementary) *use case*. The “Start the Vehicle” *use case*, is modeled as an extension of “Drive the Vehicle.” This means that there are conditions that may exist that require the execution of an instance of “Start the Vehicle” before an instance of “Drive the Vehicle” is executed.

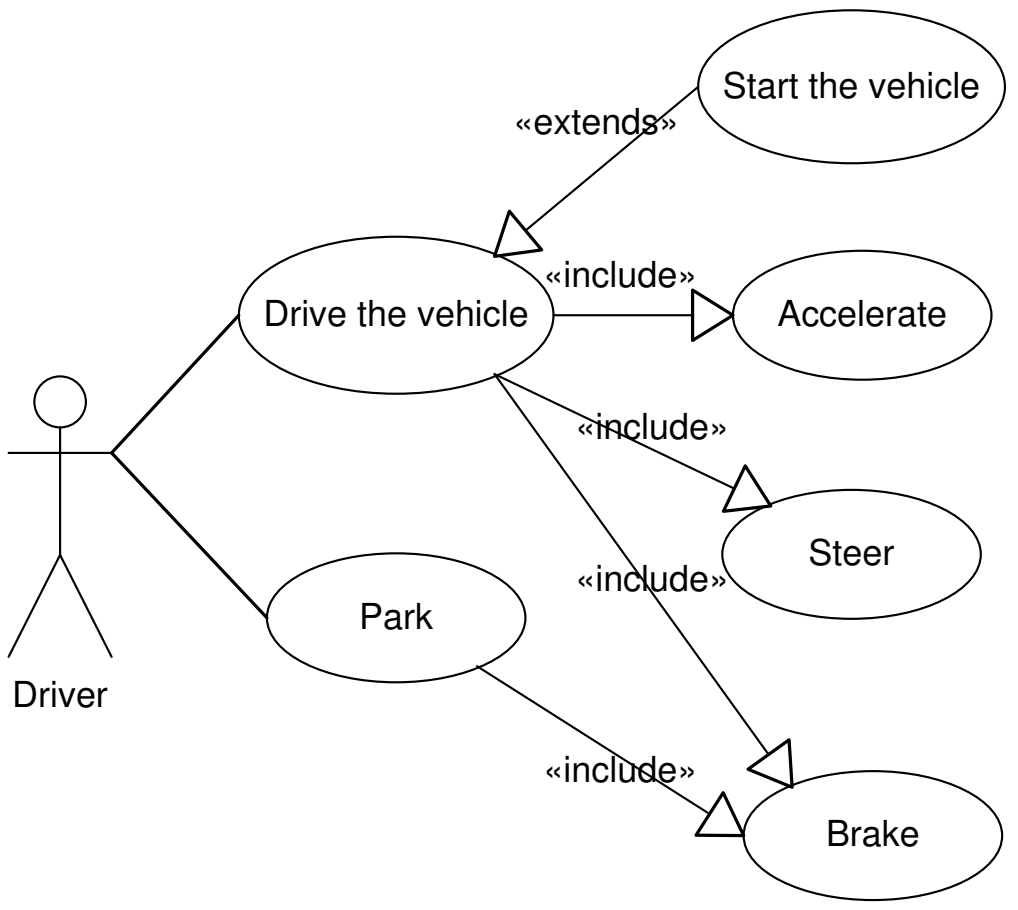

**Figure 15.** Use Case Diagram

The *use cases* “Accelerate,” “Steer,” and “Brake” are modeled using the *include* relationship. *Include* is a directed relationship between two *use cases*, implying that the behavior of the *included use case* is inserted into the behavior of the *including use case*. The *including use case* may only depend on the result (value) of the *included use case*. This value is obtained as a result of the execution of the *included use case*. This means that "Accelerate," "Steer," and "Brake" are all part of the normal process of executing an instance of "Drive the Car."

In many situations, the use of the *Include* and *Extend* relationships is subjective and may be reversed, based on the approach of an individual modeler.

With respect to DEVS-based M&S, *use case* diagrams are used to establish the system context; defining system boundaries and multi-level resolutional capabilities at appropriate hierarchical level. They should be used like a starting point in the model development. However, there are no special rules to design the 'best' *use cases*. The resolution depends mainly on the model context under development.

*4.3.2 Sequence diagrams*

The sequence diagram describes the flow of control between actors and systems or between parts of a system. This diagram represents the sending and receiving of messages between the interacting entities called lifelines, where time is represented along the vertical axis. The sequence diagrams can represent highly complex interactions with special constructs to represent various types of control logic, reference interactions on other sequence diagrams, and decomposition of lifelines into their constituent parts.

We make use of State Diagrams defined in Choi's paper [25], which uses three operator (seq, alt, and loop), and adds the DEVS "sigma" information to explicitly specify the timing constraint among components of a system. We define both the phase and sigma of the model by means of constraints in the sequence diagram. Sigma specifies a point in time and the event occurs at the specified time, the phase provides information about the model's global state. For example, the constraint "active, 5s" means that the event occurs after 5 seconds, while the global state of the model is "active". All the sending and receiving events should have sigma and phase, except those lifelines which represent coupled models. If sigma is infinity it is denoted as "inf" which implies that the object waits for the incoming event indefinitely until any message arrives. Finally, if sigma is not defined explicitly, it is supposed that sigma is updated using the elapsed time of the coming event, i.e.:

$$\sigma = \sigma - \varepsilon$$

where $\varepsilon$ is the elapsed time.

Figure 16 depicts an illustrative example. From M1 point of view, there is an initialization message which initializes phase to "active" and sigma to 5 seconds. If no external transition happens, the next event (message) is sent after 5 seconds followed by an internal transition, which sets the phase to "passive" and sigma to infinity. After 9 seconds, M1 receives a message from M2 and an external transition happens. It sets phase of M1 to passive and sigma to infinity. Note that in the case of M2, its sigma is updated to 4 seconds after the M1 internal transition (9 seconds minus elapsed time, 5 seconds) implicitly.

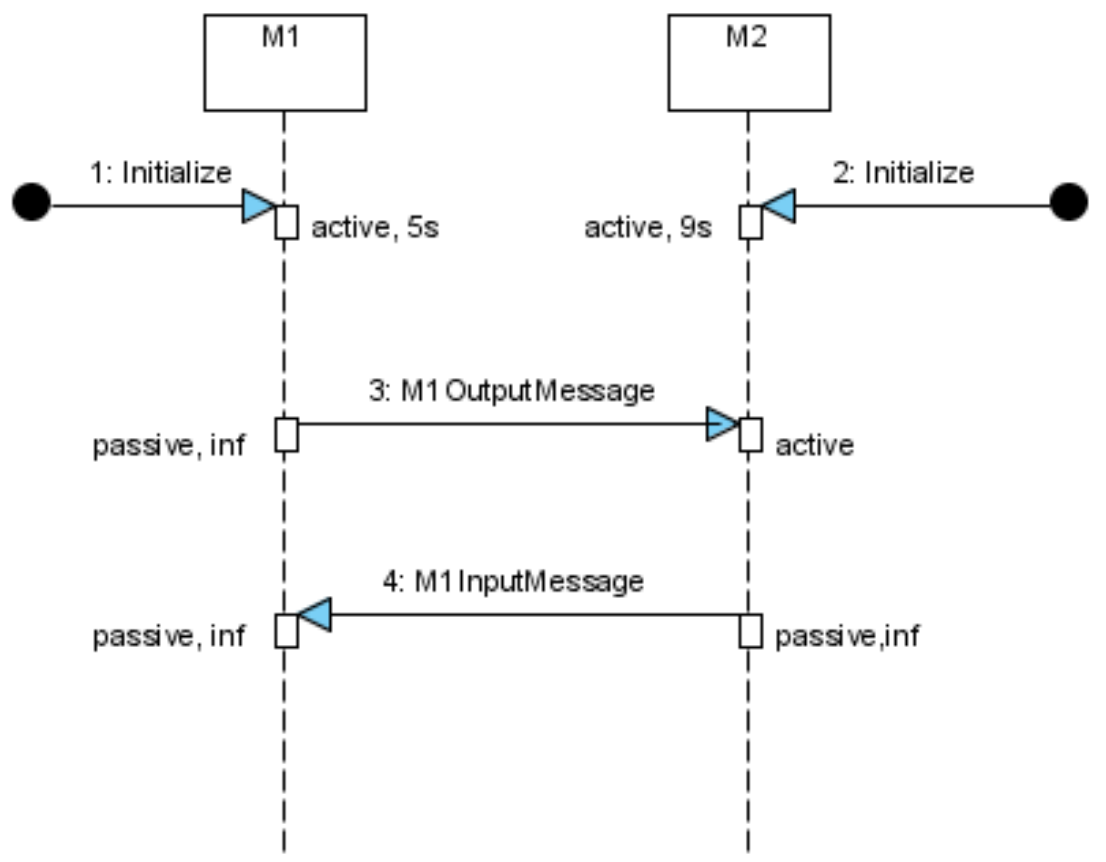

**Figure 16.** Example Sequence Diagram

Although sequence diagrams are presented in the behavior section, they are able to represent DEVS structure. Table II summarizes the relations between DEVS formalism and UML sequence diagram.

**Table II.** Relation between DEVS and UML sequence diagram

| DEVS | UML Sequence diagram |
|---|---|
| **Atomic model (Lifeline with "atomic" stereotype)** | |
| *IP* | incoming messages' name |
| *OP* | outgoing messages' name |
| *X* | incoming messages |
| *S* | constraints (phase, sigma) |
| *Y* | outgoing messages |
| **Coupled model (Lifeline with "coupled" stereotype)** | |
| *IP* | incoming messages' name |
| *OP* | outgoing messages' name |
| *X* | incoming messages |
| *Y* | outgoing messages |
| *D* | other lifelines |
| *EIC* | external incoming messages |
| *EOC* | external outgoing messages |
| *IC* | messages between lifelines |

*4.3.3 Timing diagrams*

Timing diagrams are one of the new artifacts added to UML 2. They are used to explore the behaviors of one or more objects throughout a given period of time. There are two basic flavors of timing diagram, the concise notation and the robust notation.

Figure 17 depicts an example of both concise and robust notations of timing diagrams. M1 starts in "active" state for five seconds. After that, M1 sends a *timeout message* called "M1OutputMessage" and makes an internal transition changing its state to "passive". The message changes the state of M2 from "active, 9s" to "active" (sigma=4s implicitly) through an external transition in M2. Four seconds later, M2 sends a *timeout message* and makes an internal transition, changing its state from "active" to "passive". The message is called "M1InputMessage", which executes an external transition in M1 without effects.

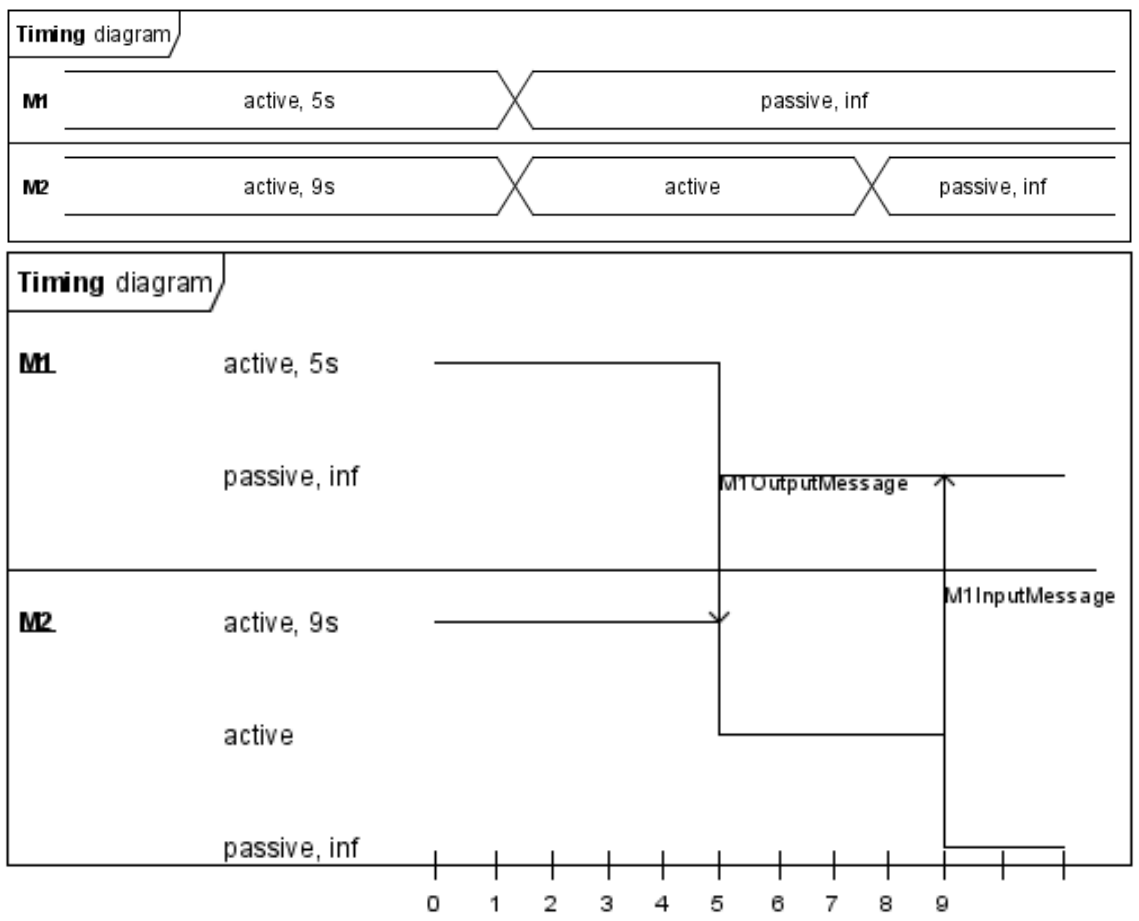

**Figure 17.** Timing diagram

Timing diagrams are valid only for DEVS atomic models. So they are able to define the models' behavior. Table III summarizes the relation between DEVS formalism and timing diagrams.

**Table III. Relation between DEVS and UML timing diagram**

| DEVS | UML Timing diagram |
|---|---|
| **Atomic model (lifeline)** | |
| *IP* | incoming time messages' name |
| *OP* | outgoing time messages' name |
| *X* | incoming time messages |
| *S* | state/condition (phase, sigma) |
| *Y* | outgoing time messages |

*4.3.4 State machine diagrams*

UML state machine diagrams define a set of concepts that can be used for modeling discrete behavior through finite state transition systems. The state machine represents behavior as the state history of an object in terms of its transitions and states. The activities that are invoked during the transition of the states are specified along with the associated event and guard conditions following the format "event[guard]/activity".

In our previous work we established a set of procedures to define a DEVS state machine by using UML state machine diagrams using SCXML [32]. We used IBM Rational Software architect to export UML state machine diagrams into SCXML and then used the XSLT mechanism (see Appendix B) to export it to XFD-DEVS. In its completeness, we showed how to transform UML state machine diagrams into DEVS executable code. In the present approach, which is guided by XFD-DEVS formalism, we follow a slightly different notation, where the state stores a list of two parameters: DEVS phase and sigma (timeout for that state). Output messages are defined by activities in the transition with the keyword *deltint* as the name of the event. It means that an internal transition happens, and just before it, an output message is sent. Input messages are specified using events in the transition, which denotes a DEVS external transition. Optionally, we utilize the guard condition to define the interface sent or received, as we show in the example section. Table IV summarizes the relation between DEVS formalism and state machine diagrams.

Figure 18 shows the state machine diagram for M1. M1 starts in state "active" for five seconds. After that the message "M1OutpurMessage" is sent, and an internal transition happens, changing the state to "(passive, inf)". If M1 receives a message "M1InputMessage", an external transition is executed, without changes in the state of M1. For more details on how a state machine is specified refer [9].

**Table IV.** Relation between DEVS and UML state machine diagram

| DEVS | UML state machine diagram |
|---|---|
| **Atomic model (diagram)** | |
| *X* | events |
| *S* | state (phase/sigma) |
| *Y* | activities |

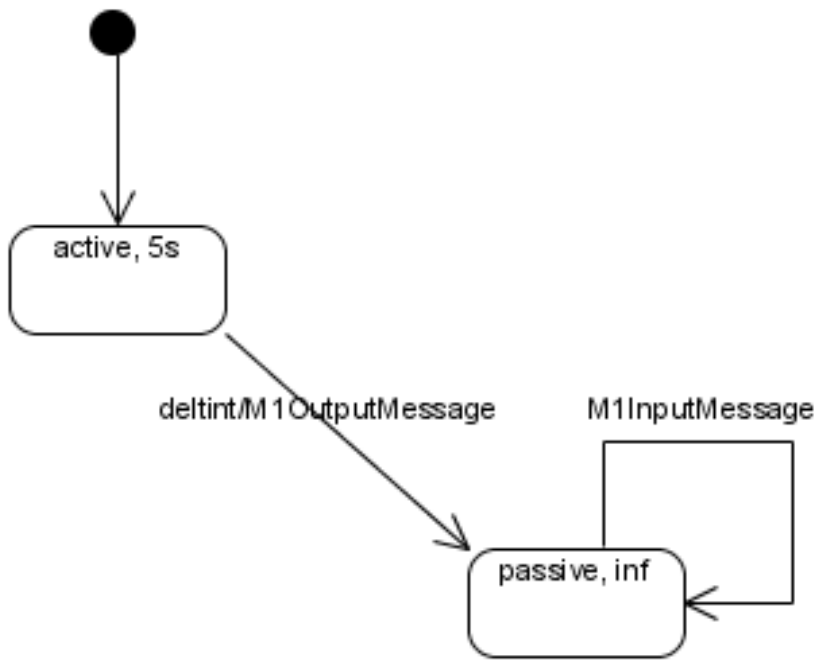

**Figure 18.** M1 state machine diagram

## 5. Transformations

In the previous section we described how to define a DEVS model (both structure and behavior) by means of seven different UML diagrams. Some of them are utilized to define the DEVS model at a high level of abstraction (external package diagrams to represent the structure and use case diagrams to represent the behavior). However, it is important to emphasize that it is sufficient to implement a UML component diagram and a UML state machine diagram to define a DEVS executable model, but in this case, the development process is closer to modeling experts than domain experts.

Modeling a DEVS model through the UML diagrams as described in earlier sections may follow a certain order. First, the structure must be defined in terms of component diagrams, package diagrams and class diagrams. This UML information can be very easily represented by SES diagram as well, which is entirely XML-based in its latest implementation. Second, the behavior is defined by means of use case diagrams and sequence diagrams or timing diagrams or state machine diagrams which are augmented by more information as per XFD-DEVS requirements.

At this stage, we have information coming from UML, SES and XFD-DEVS models. Our task is to remove redundancy and take the intersection of this information set guided by the minimalist information that is needed to create a DEVS M&S-based system. The information extraction process is largely attributed to various XML-based technologies such as XSLT, XPATH, XSD, DOM and JAXB. As laid out in Section 4.1 and bounded by Figure 10, we define our transformations leading to a DEVS PSM.

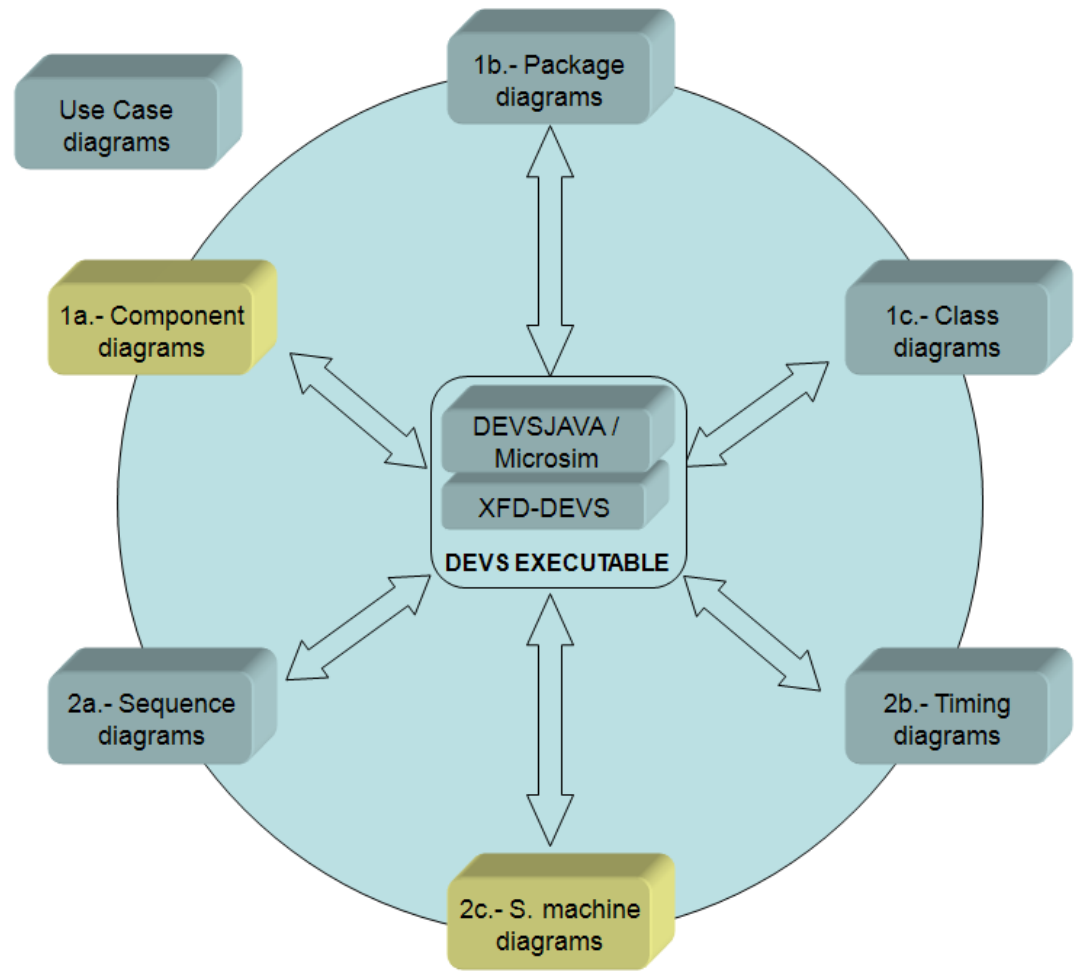

**Figure 19.** Possible transformations

Figure 19 depicts the set of possible transformations between formats (transformation between DEVSJAVA and XFD-DEVS are described in [32]). The transformation from XML files to Java files are implemented using XSLT. Transformations from DEVSJAVA files to XML files are implemented using Java XML libraries such as JavaML [37]. *Use cases* are out of the transformations because they do not provide relevant information for the DEVS executable model. In addition, component and state machine diagrams are represented in a different color because they are the minimal set to represent a DEVS executable model.

## 6. Integrating eUDEVS into DUNIP

This section describes the bifurcated Model-Continuity process [12] and how various elements like automated DEVS model generation, automated test-model generation (and net-centric simulation over SOA are put together in the process, resulting in DEVS Unified Process (DUNIP) [12], [36]. The DEVS Unified Process (DUNIP) is built on the bifurcated Model-continuity based life-cycle methodology.

The design of simulation-test framework occurs in parallel with the simulation-model of the system under design. The DUNIP process consists of the following elements:

1. Automated DEVS Model Generation from various requirement specification formats
2. Collaborative model development using DEVS Modeling Language (DEVSML)
3. Automated Generation of Test-suite from DEVS simulation model
4. Net-centric execution of model as well as test-suite over SOA

Considerable amount of effort has been spent in analyzing various forms of requirement specifications, viz., state-based, Natural Language based, UDEVS, Rule-based, BPMN/BPEL-based and DoDAF-based, and the automated processes that each one should employ to deliver DEVS hierarchical models and DEVS state machines [12], [10]. Simulation execution today is more than just model execution on a single machine. With Grid applications and collaborative computing the norm in industry as well as in scientific community, a net-centric platform using XML as middleware results in an infrastructure that supports distributed collaboration and model reuse. The infrastructure provides for a platform-free specification language DEVS Modeling Language (DEVSML) [8] and its net-centric execution using Service-Oriented Architecture called DEVS/SOA [20]. Both the DEVSML and DEVS/SOA provide distributed technologies to integrate, collaborate and remotely execute models on SOA. This infrastructure supports automated procedures in the area of test-case generation leading to test-models. Using XML as the system specifications in rule-based format, a tool known as Automated Test Case Generator (ATC-Gen) was developed which facilitated the automated development of test models [38].

DUNIP (Figure 20) can be summarized as the sequence of the following steps:

1. Develop the requirement specifications in one of the chosen formats such as BPMN, DoDAF, Natural Language Processing (NLP) based, UML based or simply DEVS-based for those who understand the DEVS formalism
2. Using the DEVS-based automated model generation process, generate the DEVS atomic and coupled models from the requirement specifications using XML
3. Validate the generated models using DEVS W3C atomic and coupled schemas to make them net-ready capable for collaborative development, if needed. This step is optional but must be executed if distributed model development is needed. The validated models which are Platform Independent Models (PIMs) in XML can participate in collaborative development using DEVSML.
4. From step 2, either the coupled model can be simulated using DEVS/SOA or a test-suite can be generated based on the DEVS models.
5. The simulation can be executed on an isolated machine or in distributed manner using SOA middleware if the focus is net-centric execution. The simulation can be executed in real-time as well as in logical time.
6. The test-suite generated from DEVS models can be executed in the same manner as laid out in Step 5.
7. The results from Step 5 and Step 6 can be compared for V&V process.

Having UML described using eUDEVS, we now have a means to incorporate UML models into DEVS based integrated modeling and simulation framework. Incorporating eUDEVS in DUNIP allows us to develop the system model along with its test suite, as laid out in the DUNIP.

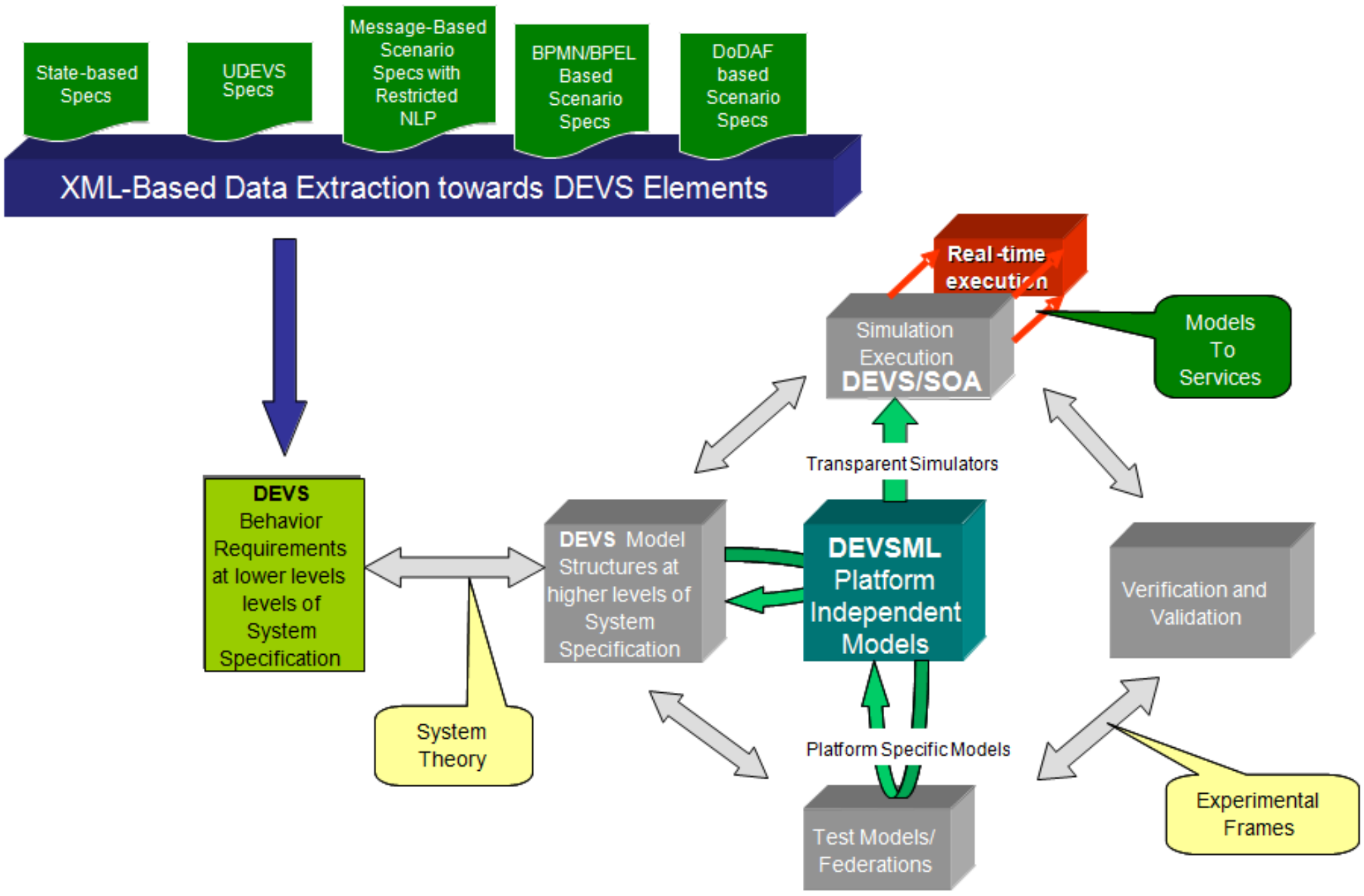

**Figure 20**: The Complete DEVS Unified Process

## 7. Case study: XFD-DEVS and UML Together

In this Section we show some of the possible cross transformations described in Section 5 between eUDEVS and XFD-DEVS. We have chosen these two paradigms because these are the most complicated and recent ones. Representing a state machine entirely in XML as a PIM is one of the major accomplishments of this research work. We implemented such transformations using Java (DOM libraries) plus a set of XSLT documents. We developed a graphical user interface called TUDEVS[1], by means of which a user is able to select the UML model as well as an XFD-DEVS model to execute the transformation in the chosen direction. The current version of TUDEVS has some limitations. The process of generating platform specific code or PSM is handled by XFD-DEVS framework i.e. the executable DEVS code is generated by XFD-DEVS so we only need to transform UML into XFD-DEVS. Transformations to package and class diagrams are not implemented. In addition, component coordinates are not generated, so the UML diagram generated needs some manual editions. We are working to improve these limitations and a prototype will soon be reported with full functionality.

[1] TUDEVS: Transformed UML and DEVS

### 7.1 The ef-p model

The ef-p model is a simple coupled model consisting of three atomic models (Figure 21).

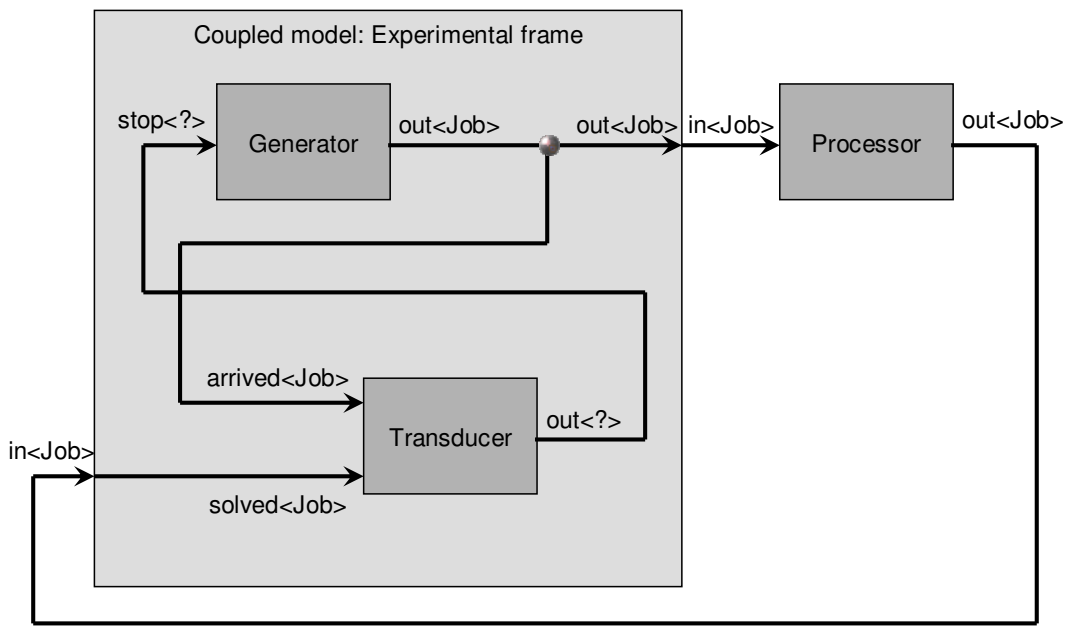

**Figure 21.** Experimental frame (ef)-processor (p) model; boxes: models; arrows: couplings; arrow labels: input/output port names.

The generator atomic model generates job-messages at fixed time intervals and sends them via the "out" port. The transducer atomic model accepts job-messages from the generator at its "arrived" port and remembers their arrival time instances. It also accepts job-messages at the "solved" port. When a message arrives at the "solved" port, the transducer matches this job with the previous job that had arrived on the "arrived" port earlier and calculates their time difference. Together, these two atomic models form an experimental frame coupled model. The experimental frame sends the generators job messages on the "out" port and forwards the messages received on its "in" port to the transducers "solved" port. The transducer observes the response (in this case the turnaround time) of messages that are injected into an observed system. The observed system in this case is the processor atomic model. A processor accepts jobs at its "in" port and sends them via "out" port again after some finite, but non-zero time period. If the processor is busy when a new job arrives, the processor discards it. Finally the transducer stops the generation of jobs by sending any event from its "out" port to the "stop" port at the generator.

### 7.2 The ef-p UML model

TUDEVS is designed to take both kinds of inputs: XFD-DEVS as well as UML models. It is a transformer that transforms in either direction. We will start with the UML design of ef-p example.

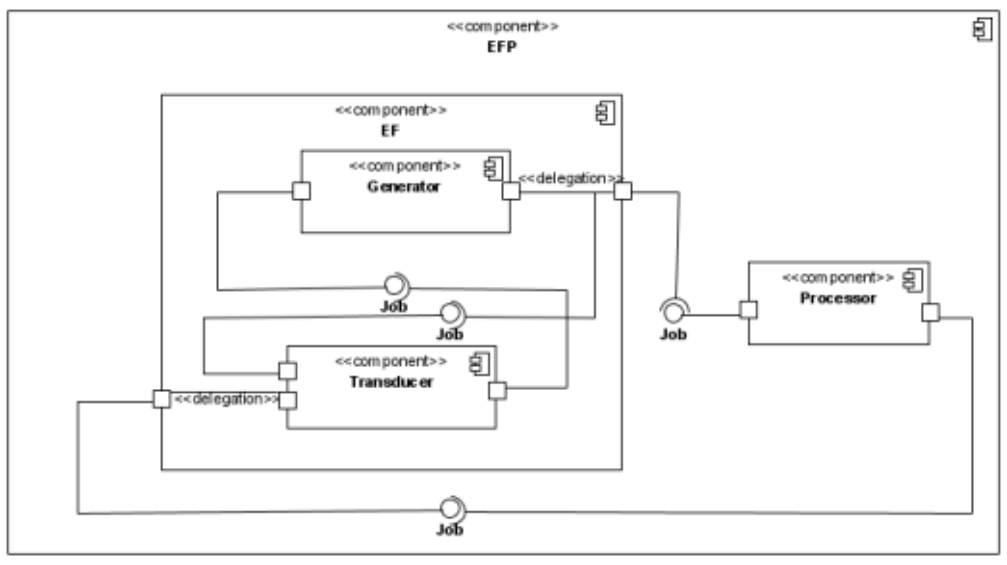

(a) Component diagram

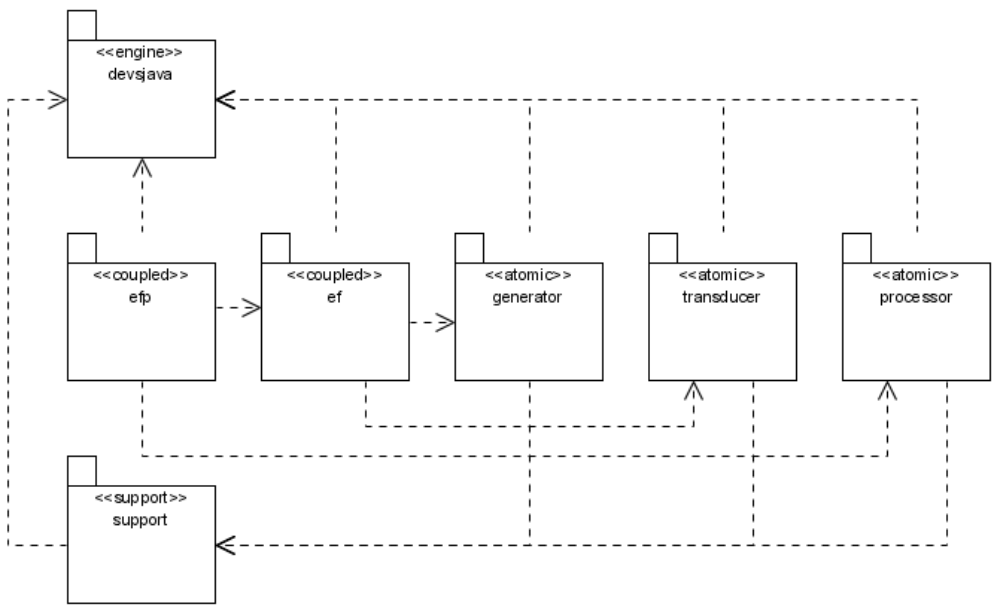

(b) Internal package diagram

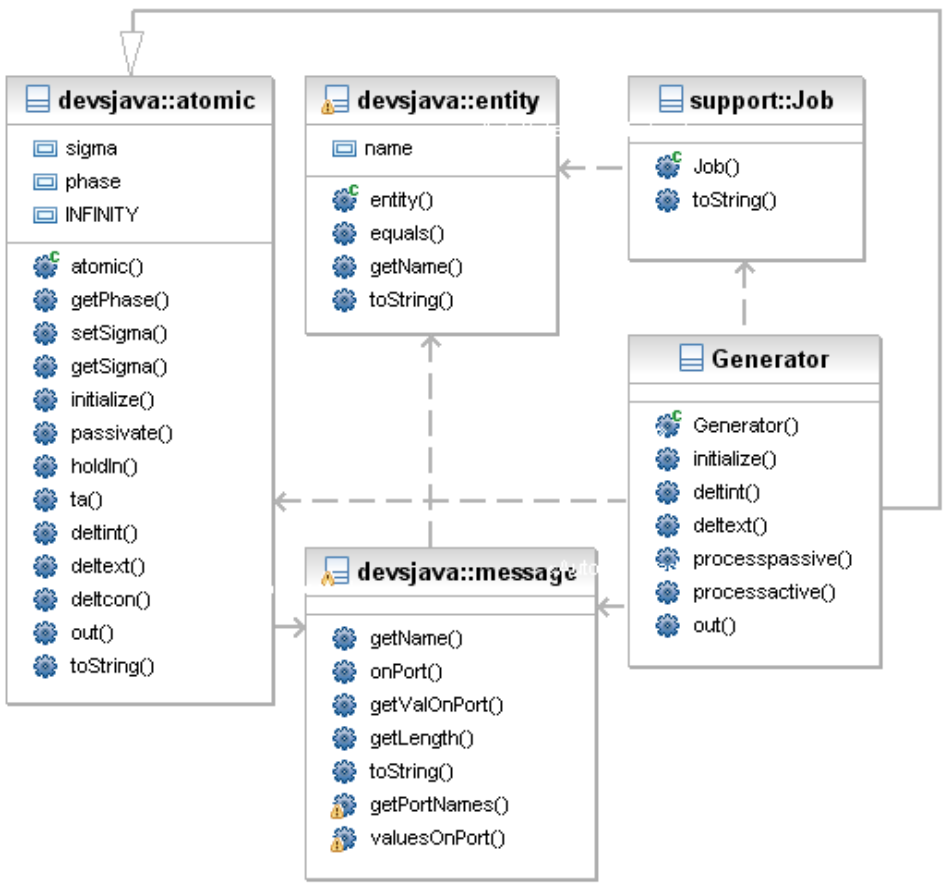

(c) Generator class diagram

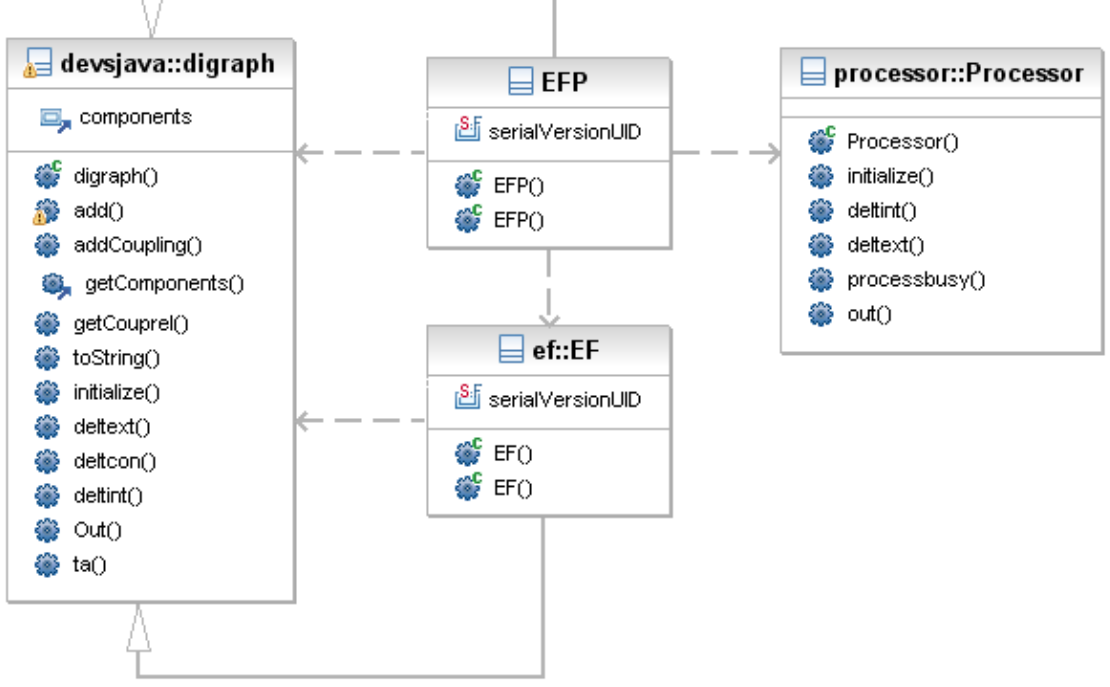

(d) ef-p class diagram

**Figure 22.** Some diagrams of the ef-p UML structure

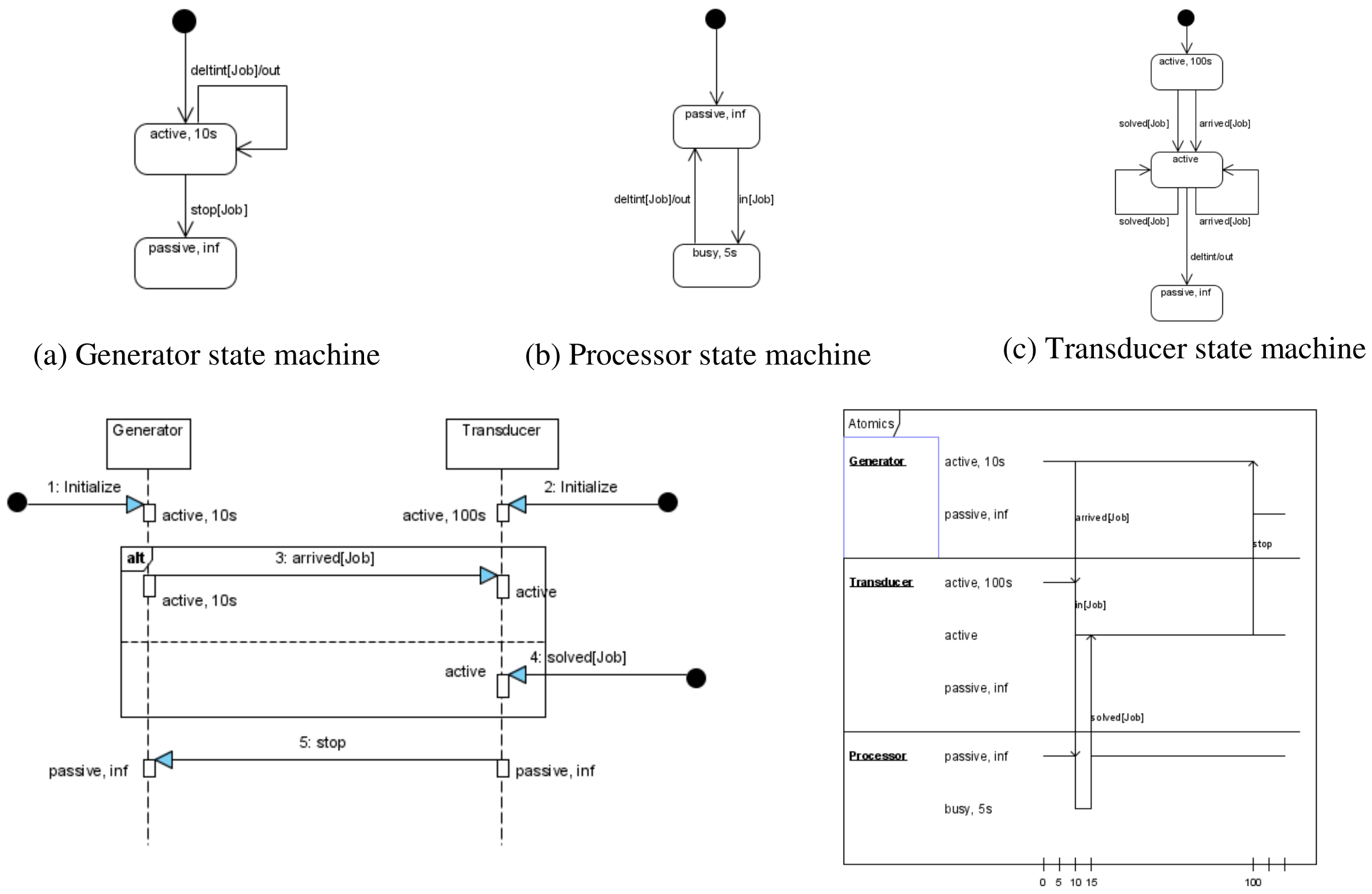

(a) Generator state machine (b) Processor state machine (c) Transducer state machine

(d) ef sequence diagram (e) Timing diagram

**Figure 23.** Some diagrams of the ef-p UML behavior

Figure 22 and 23 show the ef-p model. The structure diagrams are those shown in Figure 22a. Component diagram (Figure 22a) depicts the structure of the ef-p model in terms of components, ports, interfaces and delegation connectors. Since all the atomic models generate "Job" objects, the interface defined to create the connections between ports is precisely "Job". The internal package diagram (Figure 22b) shows the software design using the simulation engine selected, supporting classes, and components which are present in the model, such as "efp" as the root coupled model, "ef" as the experimental frame coupled model and three atomic models: "generator", "transducer" and "processor". Finally, class diagrams (Figures 22c and 22d) show the final implementations of the ef-p model. Some dependencies of DEVSJAVA have been removed for clarity reasons. In Figure 22c, the Generator class inherits "devsjava::atomic" which is the base class for atomic models in DEVSJAVA, whereas the EFP and EF classes (Figure 22d) inherit "devsjava::digraph" as the base class for coupled models. Figure 23 depicts the behavior diagrams. We did not define use cases since such diagrams are dedicated to a high-level description of the system. Figures 23a, 23b and 23c show the state machine diagrams of all the atomic models included in the ef-p. The left state machine is the generator atomic model, at the middle the processor, and finally the transducer at the right side. Figure 23d shows the sequence diagrams of the ef coupled model. Finally, Figure 23e the timing diagrams of all the atomic models.

### 7.3 From UML to XFD-DEVS

Figures 24 and 25 show how TUDEVS generates a XFD-DEVS model from UML component and state machine diagrams for the structure and behavior, respectively.

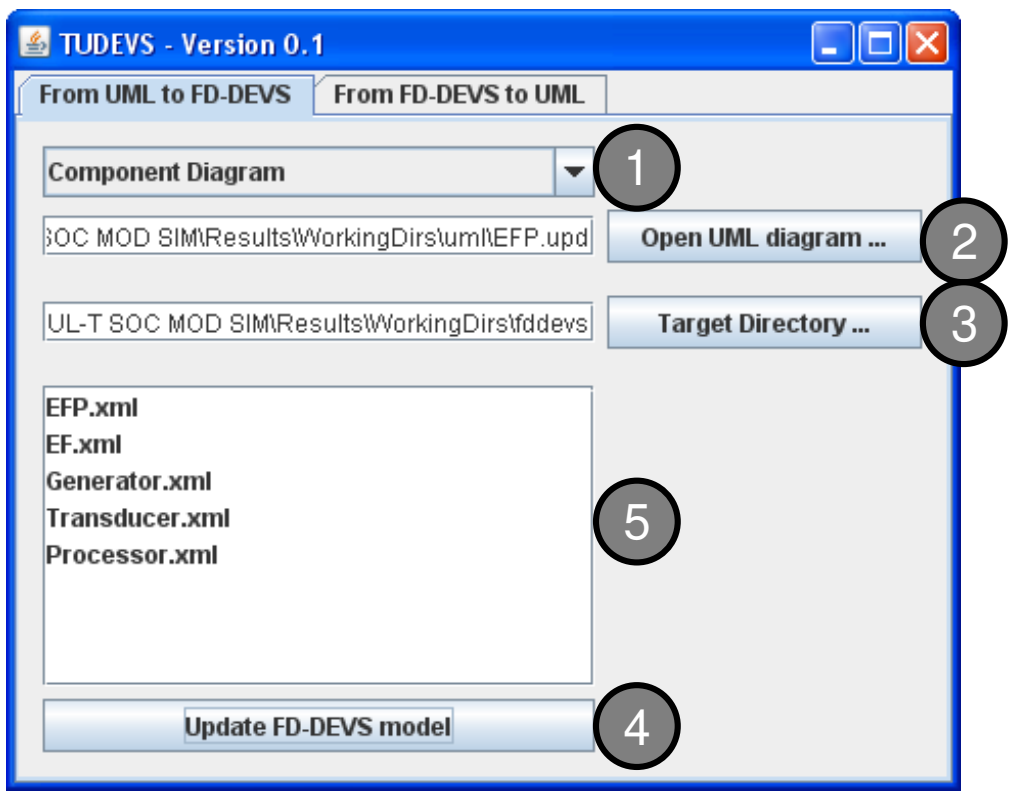

**Figure 24.** XFD-DEVS structure generation from a UML Component Diagram

First, we select the UML component diagram which will generate the corresponding XFD-DEVS structure i.e. a hierarchical coupled DEVS model. It includes all the XML files, input and output ports and connections for atomic or coupled models. Second, we select the UML XMI file. Third, we select the target directory, where the generated files will be placed. Finally, we update the current XFD-DEVS model (which is initially empty) generating the aforementioned structure.

The following XML document is the EF.xml file generated:

```
…
<Digraph>
 <Models>
  <Model>
   <devs>Generator</devs>
  </Model>
  <Model>
   <devs>Transducer</devs>
  </Model>
 </Models>
 <Couplings>
  <Coupling>
   <SrcModel>Generator</SrcModel>
   <outport>out</outport>
   <DestModel>Transducer</DestModel>
   <inport>arrived</inport>
  </Coupling>
 </Couplings>
</Digraph>
```

Next, as Figure 25 depicts, we generate the behavior of each atomic model selecting the corresponding UML State Machine diagram. In this case, the Generator XML file is updated with the behavior (states, transitions and outputs).

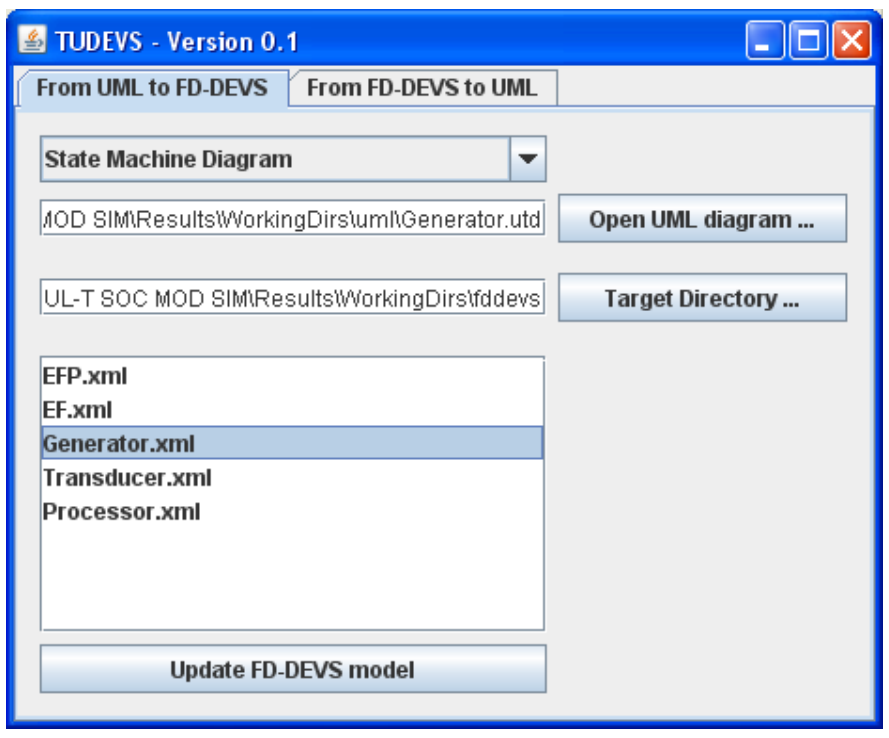

**Figure 25.** XFD-DEVS behavior generation from a UML State Machine Diagram

The following XML shows the part of the behavior section of the Generator in the XFD-DEVS model updated. The structure (input and out ports) has been omitted for brevity.

```
<Atomic>
 <states>
  <state>passive</state>
  <state>active</state>
 </states>
 <TimeAdvance>
  …
  <ta>
   <state>active</state>
   <Timeout>10.0</Timeout>
  </ta>
 </TimeAdvance>
 <LamdaSet>
  <lamda>
   <state>active</state>
   <outport>out</outport>
  </lamda>
 </LamdaSet>
 <deltint>
  <InternalTransition intTransitionID="2">
   <transition>
    <StartState>active</StartState>
    <NextState>active</NextState>
   </transition>
  </InternalTransition>
  …
 </deltint>
 <deltext>
  <ExternalTransition extTransitionID="2">
   <IncomingMessage>Job</IncomingMessage>
   <transition>
    <StartState>active</StartState>
    <NextState>passive</NextState>
   </transition>
   …
  </ExternalTransition>
  …
 </deltext>
</Atomic>
```

The complete source code of efp XFD-DEVS model is available online at [39]. Recall that this XML representation of finite state machine is completely executable using DEVSJAVA or DEVS.net or Microsim simulation framework. In the next sub section we will see how a reverse transformation from XFD-DEVS to UML can be attempted.

### 7.4 From XFD-DEVS to UML

The procedure to generate UML diagrams form XFD-DEVS models goes through the same kind of steps as in the reverse manner. Figure 26 show how we generate the component diagram. First, we select the directory where the XFD-DEVS model is located. Second, we select the XFD-DEVS files needed to generate the desired UML diagram. For example, if we want to generate a UML component diagram, we must select the source coupled model (the root coupled model in Figure 26). Next, we select the type of diagram we want to generate. Finally, we generate the UML diagram into a new XMI file. This is a work in progress and will soon be reported at [11].

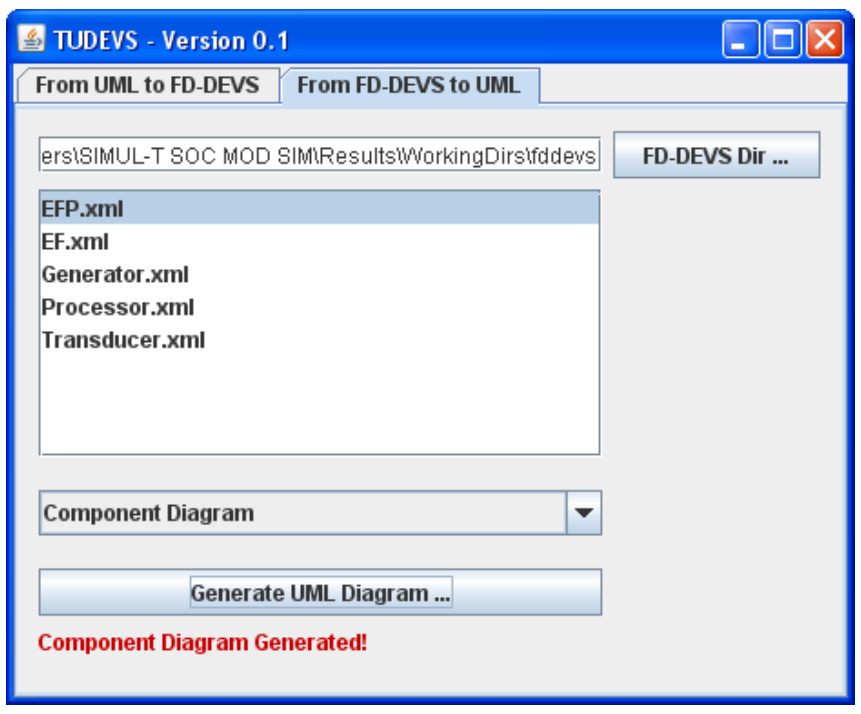

**Figure 26.** UML Component Diagram generation from the XFD-DEVS model

We have shown that how TUDEVS can take in a UML model and can generate an XFD-DEVS executable simulation model and how XFD-DEVS model can deliver a UML model (with some reservations). We are working towards removing the deficiencies in the generated UML models from TUDEVS so that they can be viewed by IDEs that have the capabilities to import/export XMI files. This example has demonstrated the proof of concept and the validation of the underlying eUDEVS metamodel that made these transformations easier.

## 8. Discussion

UML has become the defacto standard for modeling in the industry and lot of vendors provide rich graphical tools conforming to the UML 2.0 standard. There are many open-source tools such as Eclipse that propel the development of extensions related to UML additions and research. UML profiles have been defined for various application areas such as DoDAF, Systems engineering i.e. SysML and others. However, UML lacks any system theoretic foundation. DEVS on the other hand is founded on systems theoretic principles and have advocated the component based engineering since its inception. With the advancement of UML in recent years, DEVS community has made advances in mapping the DEVS elements with those of UML elements. There is a consensus that DEVS is more rigorous than UML but lacks expressive power for the non-engineer who is proficient with the graphical notations of UML. All the DEVS groups, most notably, Sarjoughian, Vangheluwe and Zinoviev have tried to address the problem of mapping DEVS and UML but without an underlying metamodel or ontology. We have proposed eUDEVS using SES ontology that binds both the UML and DEVS in a metamodel framework. This has many advantages. Having such a foundation allows eUDEVS framework to develop cross-transformations, tools and editors based on the eUDEVS metamodel. The other major advantage of such framework is inclusion of UML and its usage in a much larger systems engineering based DEVS Unified Process that allows creation of test-suite, along with the simulation model.

## 9. Conclusions and Future Work

Modeling and simulation are two independent areas. The art of modeling involves dealing with the problem domain while the simulation involves simply the execution of the model using state-of-the-art technologies and multiple platforms. Historically speaking, modeling and simulation has been integrated by the term 'simulation model' that would address any specific problem at hand. This brings a lot of problems in extending the model as well as performance of simulation itself, when both are separated. In the first place, a framework is warranted that would allow such separation.

The impact of modeling and simulation cannot be underestimated and is currently in mainstream with technologies and tools like UML, Statemate, Matlab etc. In our present research we are only dealing with software modeling domain, where UML is the preferred means and a standard. It is a graphical language that allows designers to develop their architecture model using various graphical elements. UML lies strictly in the modeling domain *as per say* and simulation of UML model, though is not a standard. There have been attempts to develop executable UML.

DEVS has been known for more than thirty years and it categorically separates the model and the simulator. However, it has not been good at the modeling interface, especially at the graphical end which accounts for its non-acceptance in the commercial non-engineering domain. The DEVS formalism exists in multiple platform specific implementations such as DEVSJAVA, DEVS/C++ and DEVS/.net. Recently a subset of DEVS, known as XFD-DEVS is made available as platform independent implementation that can lead to any of the platform specific code executions. DEVS is built on mathematical system theoretical principles so it has a very strong foundation. DEVS simulators have been made executable on P2P, RMI, CORBA, HLA and SOA that allows the same DEVS models to be simulatable on various distributed platforms as well.

In this research work we have attempted to bridge the gap between the UML graphical modeling elements and DEVS formalism by means of XML and a subset of DEVS known as XFD-DEVS that gives us the needed edge when interfacing with a powerful formalism such as DEVS. Although, this is not the first attempt, it has been the most comprehensive in terms of mapping various implementations and addressing the problem in a platform independent manner using XML based DEVS, XFD-DEVS.

This paper has made three important contributions:

1. Putting UML within a systems theoretical framework such as DEVS
2. Developing a meta-model for an executable modeling and simulation framework using ontology such as System Entity Structure (SES).
3. Making UML models executable using DEVS

Despite being graphically rich, UML suffers from lack of systems theory in the background. Extensions like SysML exist but again, they are extensions, not the needed foundation. We have developed metamodels of DEVS, UML and the proposed executable UML-Based DEVS, eUDEVS.

These metamodels are built using the SES ontology. This is by far the most important contribution where UML now is backed by systems theoretical framework.

We have described the essential mappings that need to be done to extract information from UML, or augment UML towards a DEVS model. We incorporated only the required UML elements which we thought would suffice to develop a DEVS model. These UML elements are shown as mapping to DEVS in Figure 10, which is also the metamodel for eUDEVS. We have placed eUDEVS in a much bigger framework of DEVS Unified Process (DUNIP) that is based on bifurcated model-continuity based life cycle process. This opens UML to entire suite of integrated systems development using DEVS theory. DUNIP has been applied to systems of systems engineering.

We have also demonstrated the entire life cycle of doing the cross transformation between UML and XFD-DEVS by an example that has a hierarchical component structure. We have shown how the transformations could be done with the developed tools like XFD-DEVS workbench and TUDEVS.

Finally, we have established that we have a means to develop mapping between practical frameworks (UML) with an engineering framework (DEVS) using XML transformations and SES, which lead to platform independent code. This research work closes the gap between UML and executable UML using DEVS modeling formalism and the underlying DEVS simulation protocol.

**Future Work**

We are currently in process of completing TUDEVS that allows cross transformation between XFD-DEVS and UML XMI format. The tool design will be reported in our forthcoming publication. Further there are technologies like Scalable Vector Graphics (SVG) that allow rendering XML information into graphical diagrams; we are currently incorporating this feature in TUDEVS as well. We are also exploring auto-generation of sequence diagrams form XFD-DEVS specifications using message based interfaces and temporal logic. Having the said capability of integrating UML and DEVS, we are also pursuing development of executable architectures based on Department of Defense Architecture Framework (DoDAF). Another important extension of this work is mapping with SysML for the simple reason that it is an extension of UML applied to systems engineering domain, while DEVS lie in the domain already. A mapping is underway and will be soon reported.

**Acknowledgments**

We would like to thank reviewers for providing encouraging feedback and insightful comments that improved the content and the quality of the paper.

## AUTHORS BIOGRAPHY

**José L. Risco-Martín** is an Assistant Professor in Complutense University of Madrid, Spain. He received his PhD from Complutense University of Madrid in 2004. His research interests are computational theory of modeling and simulation, with emphasis on DEVS, dynamic memory management of embedded systems, and net-centric computing. He can be reached at jlrisco@dacya.ucm.es

**Saurabh Mittal** is the CEO at DUNIP Technologies, India. Previously he worked as Research Assistant Professor at the Department of Electrical and Computer Engineering at the University of Arizona where he received his Ph. D in 2007. His areas of interest include Web-based M&S using SOA, executable architectures, distributed simulation, and System of Systems engineering using DoDAF. He can be reached at saurabh.mittal@duniptechnologies.com

**Jesús M. de la Cruz** is Professor at the Department of Computer Architecture and Automation at the Complutense University of Madrid, Spain, where he is the head of the Automatic Control and Robotics Group. His interest covers broad aspects of automatic control and its applications, real time control, simulation, optimization, statistical learning, and robotics. He can be reached at jmcruz@dacya.ucm.es

**Bernard P. Zeigler** is Professor of Electrical and Computer Engineering at the University of Arizona, Tucson and Director of the Arizona Center for Integrative Modeling and Simulation. He is developing DEVS-methodology approaches for testing mission thread end-to-end interoperability and combat effectiveness of Defense Department acquisitions and transitions to the Global Information Grid with its Service Oriented Architecture (GIG/SOA). He can be reached at zeigler@ece.arizona.edu

# APPENDIX

## A. Atomic XFD-DEVS Schema

```
<?xml version="1.0" encoding="UTF-8"?>

<xsd:schema xmlns:xsd="http://www.w3.org/2001/XMLSchema"
            targetNamespace="http://www.duniptechnologies.com/binding/devsAtomic"
            xmlns:tns="http://www.duniptechnologies.com/binding/devsAtomic"
            elementFormDefault="qualified">
    <xsd:complexType name="TransitionType">
        <xsd:sequence>
            <xsd:element name="StartState" type="xsd:string"/>
            <xsd:element name="NextState" type="xsd:string"/>
        </xsd:sequence>
    </xsd:complexType>
    <xsd:complexType name="InportsType">
        <xsd:sequence>
            <xsd:element     name="inport"     maxOccurs="unbounded"     type="xsd:string"
minOccurs="0"/>
        </xsd:sequence>
    </xsd:complexType>
    <xsd:complexType name="OutportsType">
        <xsd:sequence>
            <xsd:element      name="outport"      minOccurs="0"      maxOccurs="unbounded"
type="xsd:string"/>
        </xsd:sequence>
    </xsd:complexType>
    <xsd:complexType name="StatesType">
        <xsd:sequence>
            <xsd:element       name="state"       minOccurs="1"       maxOccurs="unbounded"
type="xsd:string"/>
        </xsd:sequence>
    </xsd:complexType>
    <xsd:complexType name="TaType">
        <xsd:sequence>
            <xsd:element name="state" type="xsd:string"/>
            <xsd:element name="timeout" type="xsd:double"/>
        </xsd:sequence>
    </xsd:complexType>
    <xsd:complexType name="TimeAdvanceType">
        <xsd:sequence>
            <xsd:element name="ta" type="tns:TaType" maxOccurs="unbounded"/>
        </xsd:sequence>
    </xsd:complexType>
    <xsd:complexType name="LamdaType">
        <xsd:sequence>
            <xsd:element name="state" type="xsd:string"/>
            <xsd:element name="outport" type="xsd:string"/>
        </xsd:sequence>
    </xsd:complexType>
    <xsd:complexType name="LamdaAllType">
        <xsd:sequence>
            <xsd:element     name="lamda"     type="tns:LamdaType"     maxOccurs="unbounded"
minOccurs="0"/>
        </xsd:sequence>
    </xsd:complexType>
    <xsd:complexType name="IntTransitionType">
        <xsd:sequence>
            <xsd:element name="transition" type="tns:TransitionType"/>
        </xsd:sequence>
        <xsd:attribute name="id" type="xsd:int" use="required"/>
    </xsd:complexType>
    <xsd:complexType name="DeltintType">
        <xsd:sequence>
            <xsd:element          name="InternalTransition"          maxOccurs="unbounded"
type="tns:IntTransitionType" minOccurs="0"/>
        </xsd:sequence>
    </xsd:complexType>
    <xsd:complexType name="ExtTransitionType">
        <xsd:sequence>
            <xsd:element name="IncomingMessage" type="xsd:string"/>
            <xsd:element name="transition" type="tns:TransitionType"/>
            <xsd:element name="ScheduleIndicator" type="xsd:boolean"/>
        </xsd:sequence>
        <xsd:attribute name="id" type="xsd:int" use="required"/>
```

```
    </xsd:complexType>
    <xsd:complexType name="DeltextType">
        <xsd:sequence>
            <xsd:element       name="ExternalTransition"       type="tns:ExtTransitionType"
maxOccurs="unbounded" minOccurs="0"/>
        </xsd:sequence>
    </xsd:complexType>
    <xsd:element name="Atomic">
        <xsd:complexType>
            <xsd:sequence>
                <xsd:element name="inports" type="tns:InportsType"/>
                <xsd:element name="states" type="tns:StatesType"/>
                <xsd:element name="outports" type="tns:OutportsType"/>
                <xsd:element name="deltint" type="tns:DeltintType"/>
                <xsd:element name="deltext" type="tns:DeltextType"/>
                <xsd:element name="timeAdvance" type="tns:TimeAdvanceType"/>
                <xsd:element name="lamdas" type="tns:LamdaAllType"/>
            </xsd:sequence>
            <xsd:attribute name="modelName" type="xsd:string" use="required"/>
            <xsd:attribute name="host" type="xsd:string" use="required"/>
        </xsd:complexType>
    </xsd:element>
</xsd:schema>
```

## B. SCXML to XFD-DEVS XSLT description

```
<?xml version="1.0" encoding="UTF-8"?>

<xsl:stylesheet version="1.0" xmlns:xsl="http://www.w3.org/1999/XSL/Transform">
    <xsl:output method="xml" indent="yes"/>
    <xsl:template match="/scxml">
            <statemachine name="default" host="localhost">
                    <!-- First, deltint function -->
                    <xsl:if test="count(//send)>0">
                            <deltint>
                                    <transitionsInt>
                                            <xsl:apply-templates select="//send"
mode="deltint"/>
                                    </transitionsInt>
                            </deltint>
                    </xsl:if>
                    <!-- Second, deltext function -->
                    <deltext>
                            <transitionsExt>
                                    <xsl:apply-templates
select="//transition[count(@event)>0]" mode="deltext"/>
                            </transitionsExt>
                    </deltext>
            </statemachine>
    </xsl:template>

    <xsl:template match="send" mode="deltint">
            <xsl:variable name="event" select="@event"/>
            <!-- send is processed independly if the parent -->
            <!-- is onentry, onexit, transition or finalize -->
            <transition>
                    <startState><xsl:value-of select="../../@id"/></startState>
                    <!-- To obtain the next state, I try to find the transition
corresponding to this event -->
                    <xsl:apply-templates select="//transition[@event=$event]"
mode="deltint"/>
                    <!-- I suppose delay is always given -->
                    <timeout><xsl:value-of select="@delay"/></timeout>
                    <!-- I suppose event is always given -->
                    <outMsg><xsl:value-of select="@event"/></outMsg>
            </transition>
    </xsl:template>

    <xsl:template match="transition" mode="deltint">
            <nextState><xsl:value-of select="@target"/></nextState>
    </xsl:template>

    <xsl:template match="transition" mode="deltext">
            <xsl:variable name="event" select="@event"/>
            <xsl:if test="count(//send[@event=$event])=0">
```

```
                    <transitionExt>
                        <incomingMsg><xsl:value-of select="@event"/></incomingMsg>
                        <transition>
                            <startState><xsl:value-of select="../@id"/></startState>
                            <nextState><xsl:value-of select="@target"/></nextState>
                            <timeout>0</timeout>
                            <outMsg><xsl:value-of select="@event"/></outMsg>
                        </transition>
                    </transitionExt>
                </xsl:if>
        </xsl:template>

</xsl:stylesheet>
```

## C. Coupled XFD-DEVS Schema

```
<?xml version="1.0" encoding="UTF-8"?>

<xsd:schema xmlns:xsd="http://www.w3.org/2001/XMLSchema"
            targetNamespace="http://www.duniptechnologies.com/binding/devsCoupled"
            xmlns:tns="http://www.duniptechnologies.com/binding/devsCoupled"
            elementFormDefault="qualified">
    <xsd:element name="Digraph">
        <xsd:complexType>
            <xsd:sequence>
                <xsd:element name="Couplings" type="tns:CouplingsType"/>
                <xsd:element name="Models" type="tns:ModelsType"/>
                <xsd:element name="Inports" type="tns:InportsType"/>
                <xsd:element name="Outports" type="tns:OutportsType"/>
            </xsd:sequence>
            <xsd:attribute name="name" type="xsd:string"/>
            <xsd:attribute name="host" type="xsd:string"/>
        </xsd:complexType>
    </xsd:element>
    <xsd:complexType name="CouplingType">
        <xsd:sequence>
            <xsd:element name="src">
                <xsd:simpleType>
                    <xsd:restriction base="xsd:string"/>
                </xsd:simpleType>
            </xsd:element>
            <xsd:element name="dest">
                <xsd:simpleType>
                    <xsd:restriction base="xsd:string"/>
                </xsd:simpleType>
            </xsd:element>
            <xsd:element name="outport">
                <xsd:simpleType>
                    <xsd:restriction base="xsd:string"/>
                </xsd:simpleType>
            </xsd:element>
            <xsd:element name="inport">
                <xsd:simpleType>
                    <xsd:restriction base="xsd:string"/>
                </xsd:simpleType>
            </xsd:element>
        </xsd:sequence>
    </xsd:complexType>
    <xsd:complexType name="CouplingsType">
        <xsd:sequence>
            <xsd:element name="coupling" maxOccurs="unbounded" type="tns:CouplingType"/>
        </xsd:sequence>
    </xsd:complexType>
    <xsd:complexType name="ModelsType">
        <xsd:sequence>
            <xsd:element name="Model" maxOccurs="unbounded">
                <xsd:complexType>
                    <xsd:simpleContent>
                        <xsd:extension base="xsd:string">
                            <xsd:attribute name="type" type="xsd:string"/>
                            <xsd:attribute name="platform" type="xsd:string"/>
                        </xsd:extension>
                    </xsd:simpleContent>
                </xsd:complexType>
```

```
            </xsd:element>
        </xsd:sequence>
    </xsd:complexType>
    <xsd:complexType name="InportsType">
        <xsd:sequence>
            <xsd:element     name="inport"     type="xsd:string"     maxOccurs="unbounded"
minOccurs="0"/>
        </xsd:sequence>
    </xsd:complexType>
    <xsd:complexType name="OutportsType">
        <xsd:sequence>
            <xsd:element       name="outport"       type="xsd:string"       minOccurs="0"
maxOccurs="unbounded"/>
        </xsd:sequence>
    </xsd:complexType>
</xsd:schema>
```